\documentclass[10pt,journal]{IEEEtran}

\IEEEoverridecommandlockouts

\def\boxend{\hspace*{\fill} $\QED$}
\newtheorem{lemma}{Lemma}
\newtheorem{theorem}{Theorem}
\newtheorem{corollary}{Corollary}

\newtheorem{definition}{Definition}

\newtheorem{remark}{Remark}
\def\done{\boxend}

\usepackage{cite}
\usepackage{amsmath}
\usepackage{amssymb}
\usepackage{graphicx}
\usepackage{subfigure}
\usepackage{array}
\usepackage{hyperref}
\usepackage{multirow}
\usepackage{color}
\usepackage{enumerate}

\begin{document}
\bibliographystyle{IEEEtran}

\title{Connectivity of Underlay Cognitive Radio Networks with Directional Antennas}


\author{Qiu Wang, Hong-Ning Dai, Orestis Georgiou, Zhiguo Shi, Wei Zhang
\thanks{Q. Wang and H.-N. Dai are with Faculty of Information Technology, Macau University of Science and Technology, Macau SAR (email: qiu$\_$wang@foxmail.com; hndai@ieee.org).}
\thanks{O. Georgiou is with Ultrahaptics and the University of Bristol, Bristol, United Kingdom (email: orestis.georgiou@gmail.com).}
\thanks{Z. Shi is with College of Information Science \& Electronic Engineering, Zhejiang University, Hangzhou, P. R. China (email: shizg@zju.edu.cn).}
\thanks{W. Zhang is with School of Electrical Engineering \& Telecommunications, The University of New South Wales, Sydney, NSW 2052, Australia (email: wzhang@ee.unsw.edu.au).}
}

\maketitle

\begin{abstract}
In cognitive radio networks (CRNs), the connectivity of secondary users (SUs) is difficult to be guaranteed due to the existence of primary users (PUs). Most prior studies only consider cognitive radio networks equipped with omni-directional antennas causing \emph{high interference} at SUs. We name such CRNs with omni-directional antennas as Omn-CRNs. Compared with an omni-directional antenna, a directional antenna can concentrate the transmitting/receiving capability at a certain direction, consequently resulting in less interference. In this paper, we investigate the connectivity of SUs in CRNs with directional antennas (named as Dir-CRNs). In particular, we derive closed-form expressions of the connectivity of SUs of both Dir-CRNs and Omn-CRNs, thus enabling tractability. We show that the connectivity of SUs is mainly affected by two constraints: the \emph{spectrum availability} of SUs and the \emph{topological connectivity} of SUs. Extensive simulations validate the accuracy of our proposed models. Meanwhile, we also show that Dir-CRNs can have higher connectivity than Omn-CRNs mainly due to the lower interference, the higher spectrum availability and the higher topological connectivity brought by directional antennas.
\end{abstract}


\section{Introduction}
Cognitive radio is a promising technology to improve the efficiency of spectrum usage and to meet the growing demands of high-speed data communications \cite{Ahmad:survey2015}. Cognitive radio networks (CRNs) allow unlicenced secondary users (SUs) to opportunistically access to the spectrum without hampering the communications of licenced primary users (PUs) \cite{HSun:survey2013}. CRNs can be roughly categorized into \emph{overlay}, \emph{interweave} and \emph{underlay} paradigms. Overlay CRNs are relatively difficult to be implemented due to the prerequisite of the prior information of PUs \cite{Sibomana:TMC16}. The implementation of the interweave scheme is also challenging because it requires the perfect detection of the existence of PUs, which however is difficult to be implemented in practice \cite{Kaushik:TCCN16}. In an underlay CRN, both SUs and PUs can concurrently use the same spectrum under the provision that the interference caused by SUs at the primary receiver is below a predefined threshold \cite{Sibomana:TMC16}. As a result, a more efficient spectrum utilization can be achieved in underlay CRNs compared with overlay and interweave approaches. Therefore, in this paper we mainly consider underlay CRNs.

\subsection{Related Work and Motivation}

\begin{figure}[t]
\centering
\subfigure[Omn-CRNs]{
\begin{minipage}{3.7cm}
\centering
\includegraphics[width=3.2cm]{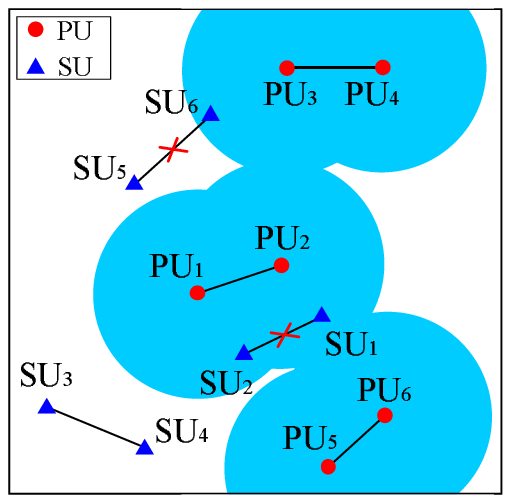}
\end{minipage}
\label{subfig:omn}
}
\subfigure[Dir-CRNs]{
\begin{minipage}{3.7cm}
\centering
\includegraphics[width=3.2cm]{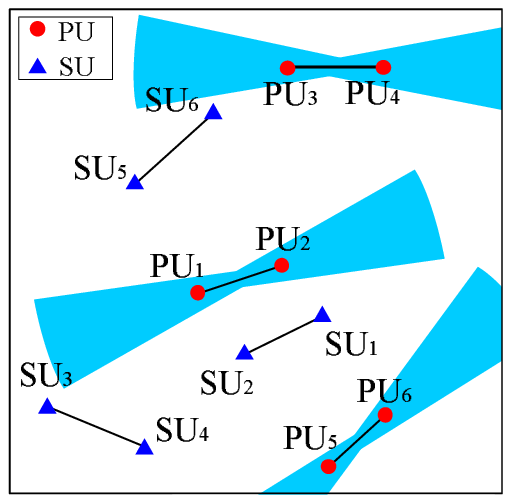}
\end{minipage}
\label{subfig:dir}
}
\caption{Omn-CRNs versus Dir-CRNs}
\label{fig:omn_dir}
\vspace*{-0.5cm}
\end{figure}

In underlay CRNs, connectivity is an important property depicting whether two nodes can establish a communication link. The connectivity of SUs in CRNs is more difficult to be ensured than that of PUs since SUs are susceptible to the existence of PUs having the higher priority to access to the spectrum than SUs. In this paper, we mainly consider the connectivity of SUs. Recently, a number of studies \cite{LiuDong:2014,YushanLiu:2015,Liu:adhoc2016} concentrate on analyzing the connectivity and delay of underlay CRNs. However, most of the studies only consider equipping both PUs and SUs with omni-directional antennas that can cause high interference due to the effect of radiating/receiving signal equally in all directions. We call such CRNs equipped with omni-directional antennas only as \emph{Omn-CRNs}.  Take Fig. \ref{subfig:omn} as an example. In this Omn-CRN, only one pair of SUs (i.e., SU$_3$ and SU$_4$) can establish a communication link while other SUs such as SU$_1$, SU$_2$, SU$_5$ and SU$_6$ cannot be active due to the existence of nearby PUs (note that the transmission region of a PU is represented by a shaded circle).

Different from omni-directional antennas, directional antennas can concentrate radio signals on desired directions and consequently reduce the interference. Thus, using directional antennas in CRNs can significantly reduce the interference as indicated in \cite{Guo:WCM2015,Dai:2015,Dung:2016}. However, most recent studies only consider using directional antennas at either PUs \cite{Dai:2015} or SUs \cite{Guo:WCM2015,Dung:2016,Yazdani:ICASSP2017} but not both. The partial deployment of directional antennas in CRNs cannot fully realise the benefits of directional antennas. Therefore, in this paper we propose a novel CRN, in which both PUs and SUs are equipped with directional antennas. We name such CRNs equipped with directional antennas as Dir-CRNs. In a nutshell, Dir-CRNs have the following characteristics: (1) each PU is equipped with a directional antenna; (2) each SU is equipped with a directional antenna; (3) SUs can use the same spectrum as PUs only when the interference caused by SUs is less than a given threshold. Note that we also consider CRNs with SUs equipped with directional antennas for comparison purpose in this paper. We name such CRNs with PUs equipped with omni-directional antennas and SUs equipped with directional antennas as Omn-Dir-CRNs.

\emph{Dir-CRNs can potentially improve the connectivity of SUs compared with Omn-CRNs.} Take Fig. \ref{subfig:dir} as an example where we consider the same placement of nodes as that in Fig. \ref{subfig:omn} for comparison. In contrast to the Omn-CRN as shown in Fig. \ref{subfig:omn}, there are 3 pairs of SUs (i.e., SU$_3$ and SU$_4$, SU$_1$ and SU$_1$, SU$_5$ and SU$_6$) that can establish the communication links under the same placement of nodes (note that the transmission range of PUs in Dir-CRNs is longer than that in Omn-CRNs due to the higher antenna gains of directional antennas). This is mainly due to the improvement of spectrum reuse and the inference reduction of directional antennas.

The goal of this paper is to investigate the connectivity of SUs in Dir-CRNs. \emph{To the best of our knowledge, there is no study on the connectivity of SUs in Dir-CRNs.} In particular, it is non-trivial to analyse the connectivity of SUs in Dir-CRNs because the connectivity of SUs in Dir-CRNs depends on multiple factors, such as the \emph{spectrum availability}, the \emph{topological connectivity}, the channel condition and the \emph{directivity} of antennas.

\subsection{Main Contributions}
This paper aims to investigate the connectivity of SUs in both Omn-CRNs and Dir-CRNs. The primary research contributions of this paper can be summarized as follows.
\begin{enumerate}[(1)]
\item We formally identify Dir-CRNs that characterize the features of equipping both PUs and SUs with directional antennas.
\item We establish a theoretical model to analyse the connectivity of SUs in both Dir-CRNs and Omn-CRNs. In particular, we derive closed-form expressions of the connectivity of SUs in both Dir-CRNs and Omn-CRNs.
\item Extensive simulations validate the accuracy of our proposed model. Specifically, our simulations results show that our analytical model can accurately analyse the connectivity of SUs in both Dir-CRNs and Omn-CRNs.
\end{enumerate}

The rest of the paper is organized as follows. Section \ref{sec:system} presents system models. We then analyse
the connectivity in Section \ref{sec:connectivity}. Section \ref{sec:sim} presents the simulation results. Finally, the paper is concluded in Section \ref{sec:conclusion}.

\section{System Models}
\label{sec:system}

\subsection{Network Model}
\label{Network Models}

In this paper, we mainly consider two types of CRNs: 1) Omn-CRNs in which both PUs and SUs are equipped with omni-directional antennas; 2) Dir-CRNs in which both PUs and SUs are equipped with directional antennas. Omn-CRNs correspond to conventional cognitive radio wireless networks in which PUs are usually referred to base stations (macro-cell base stations) or user equipments (UEs) and SUs are UEs \cite{MNi:JSAC15}. Recently, there is a new trend of using millimeter-wave (mmWave) bands in wireless networks to achieve the extremely high throughput \cite{JQiao:IEEECommMag15,Deng:ComMag17}. It becomes feasible to equip both BSs and UEs with directional antennas in mmWave CRNs since the antenna can be quite compact (as the antenna size is inversely proportional to the radio frequency). This type of mmWave CRNs corresponds to the proposed Dir-CRNs. Note that we also consider CRNs with SUs equipped with directional antennas and PUs equipped with omni-directional antennas for comparison purpose. We name this kind of CRNs as Omn-Dir-CRNs, which have been proposed in \cite{Guo:WCM2015,Yazdani:ICASSP2017}.

In a CRN (either Omn-CRN, Dir-CRN or Omn-Dir-CRN), primary transmitters (PTs) are distributed according to homogeneous Possion point process (PPP) with density $\lambda_p$ in an infinite two-dimensional Euclidean space. Each PT is associated with a primary receiver (PR), which is randomly and uniformly distributed in the transmission region of PT. According to the displacement theorem \cite{Kingman:1993}, the distribution of PRs also follows homogeneous PPP with the same density $\lambda_p$. Similarly, secondary users (i.e., SUs including both STs and SRs) are also distributed according to homogeneous PPP with density $\lambda_s$. We also assume that SUs are sparsely distributed in the network and there is no overlapping in the communication regions of any two SUs.

\subsection{Channel Model}

The radio signal is assumed to undergo both the path loss attenuation and Rayleigh fading \cite{Andrews:2011,Georgiou:2015}. In particular, the path loss attenuation is characterized by the path loss exponent $\alpha$ (usually $2 \leq \alpha \leq 6$ \cite{Rappaport:2002}). Rayleigh fading is modeled as a random variable $h$ following an exponential distribution with mean $1$. Let $P_t$ represent the transmitting power. Then, the received power denoted by $P_r$ at a receiver can be expressed as
\begin{equation}
\small
\label{eq:pr}
\small
P_r=P_{t}r^{-\alpha }h G_tG_r,
\end{equation}
where $r$ is the Euclidean distance between the transmitter and the receiver and $G_t$ and $G_r$ are the antenna gains of the transmitter and the receiver, respectively. We next describe the antenna models as well as antenna gains.

\subsection{Directional Antennas}
\label{sec:directional}

\begin{figure}[t]
\centering
\subfigure[Realistic model]{
\begin{minipage}{4.2cm}
\centering
\includegraphics[width=3.9cm]{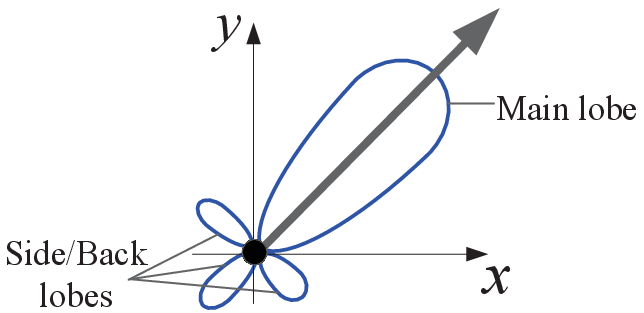}
\end{minipage}
\label{fig:real}
}
\subfigure[Sector model]{
\begin{minipage}{3.5cm}
\centering
\includegraphics[width=2.2cm]{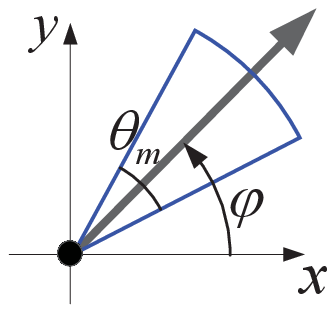}
\end{minipage}
\label{fig:sector}
}
\caption{Directional antennas models}
\vspace*{-0.5cm}
\end{figure}

We introduce the \emph{antenna gain} to measure the \emph{directivity} of an antenna. We denote the antenna gain of an omni-directional antenna by $G_o$. It is obvious that $G_o=1$ since an omni-directional antenna radiates/receives radio signals uniformly in all directions. Different from omni-directional antennas, directional antennas can concentrate transmitting or receiving capability on desired directions. Generally, a realistic directional antenna typically consists of one or several \emph{main} beams with the maximum gain and a number of side/back-lobes with the relatively lower gains, as shown in Fig. \ref{fig:real}. However, the realistic antenna models are so complicated that they are not tractable in theoretic analysis \cite{hndai2013:WPC,Renzo2015,Georgiou:2015}. A sector model is one of typical simplified antenna models \cite{zhang2010capacity,Nitsche:IEEEComMag14,hndai:jwcn15}, as shown in Fig. \ref{fig:sector}. A sector model consists of one main beam with beamwidth $\theta_m$ and all the side/back lobes are ignored. The antenna gain $G_d$ of a sector model is given as follows \cite{zhang2010capacity,Nitsche:IEEEComMag14}:
\begin{equation}
\small
G_d(\theta)=\left\{
\begin{aligned}
&\frac{2\pi}{\theta_m} & \theta\in\left(\varphi-\frac{\theta_m}{2}, \varphi+\frac{\theta_m}{2}\right), \\
&0 & \mathrm{others}, \\
\end{aligned}
\right.
\label{eq:G_sector}
\end{equation}
where $\theta\in[0,2\pi]$ is the angle from the $x$-axis in the 2-D coordinate system and $\varphi$ is the angle of the antenna orientation from $x$-axis. Note that we usually have the main beam $\theta_m<\pi$.

Eq. (\ref{eq:G_sector}) is a general expression of the antenna gain $G_d(\theta)$, which mainly depends on the beamwidth $\theta_m$. We consider this sector antenna model in our Dir-CRNs. We denote the antenna beamwidth of PUs and that of SUs by $\theta_p$ and $\theta_s$, respectively, and denote the antenna gain of PUs and that of SUs by $G_p$ and $G_s$, respectively. Note that $\theta_p$ is not necessarily equal to $\theta_s$. Replacing $\theta_m$ in Eq. (\ref{eq:G_sector}) by $\theta_p$ and $\theta_s$, we then have $G_p=\frac{2\pi}{\theta_p}$ and $G_s=\frac{2\pi}{\theta_s}$. Moreover, we denote the antenna orientation of PTs and that of STs by $\varphi_p$ and $\varphi_s$, respectively. We assume that both $\varphi_p$ and $\varphi_s$ follow the uniformly independent identical distribution (i.i.d.) within $[0,2\pi]$. Once the antenna orientation of a PT or an ST is determined, the corresponding PR or SR can adjust its antenna orientation towards the PT or the ST according to the beam-locking schemes \cite{Takai:Martin,Balanis:2005}.

\subsection{Interference constraint}
\label{sec:int_contraint}

In the underlay spectrum sharing scheme of CRNs, SUs can only access to the spectrum when the interference to PUs (either PTs or PRs) is below an acceptable threshold. In this paper, we implement a \emph{detect-and-avoid} protocol proposed in \cite{Ghasemi:ComMag08,Rabbachin:2011}. This protocol claims that each PR first transmits a detection preamble. If the received power of the preamble at an SU is greater than a threshold $\eta$, the SU becomes silent (i.e., it cannot access to the spectrum). In other words, an SU cannot have spectrum if the following condition is satisfied,
\begin{equation}
\small
P_{d}G_pG_s h R_{ps}^{-\alpha}>\eta,
\label{eq:SINR_d}
\end{equation}
where $P_{d}$ is the power of detection preamble of PRs (also called the detection power in short), $G_p$ and $G_s$ are the antenna gains of PUs and SUs, respectively, and $R_{ps}$ denotes the distance between a PU and an SU. There are different cases to determine $R_{ps}$ with regard to Dir-CRNs and Omn-CRNs (details will be given in Section \ref{subsec:spectr_avail}). In this manner, SUs can effectively avoid the interference to PUs.

\section{Connectivity}
\label{sec:connectivity}



We first define a metric of the connectivity of SUs as the \emph{probability of connection}, denoted by $p_{con}$, as follows.
\begin{definition}
\emph{Probability of connection} is the probability that any SU pair can successfully establish a bidirectional link.
\end{definition}

We then denote the probability of connection of any SU pair in Dir-CRNs, that in Omn-CRNs and that in Omn-Dir-CRNs by $p^d_{con}$, $p^o_{con}$ and $p^{od}_{con}$, respectively. Note that we consider bidirectional links since they can guarantee delivering the acknowledgement successfully (e.g., ACK in Wi-Fi). In conventional wireless networks, two nodes can establish a link if they fall into the communication region of each other. We define the condition that any two SUs fall into the communication region of each other\footnote{There is an extra condition in Omn-Dir-CRNs and Dir-CRNs: two SUs point their antennas toward each other.} as the \emph{topologically-connected condition}. Different from conventional wireless networks, the condition that two SUs in CRNs can establish a link depends on not only the topologically-connected condition \cite{Liu:adhoc2016} but also the condition that the spectrum is available to SUs. We define the condition that the spectrum is available as the \emph{spectrum availability}.

We next define the link condition of SUs as follows.
\begin{definition}
\emph{Link Condition} of SUs. An SU pair can successfully establish a link if and only if both the following conditions are satisfied:
\begin{enumerate}
\item SU pair has spectrum available;
\item SU pair is topologically connected with each other.
\end{enumerate}
\end{definition}

We denote the event that an SU pair has spectrum available by $e_{spe}$ and the event that an SU pair is topologically connected by $e_{top}$. It follows that the probability of connection $p_{con}=p(e_{spe}  e_{top})=p(e_{spe})p(e_{top}|e_{spe})$, where $p(e_{spe})$ is the probability that an SU pair has spectrum available and $p(e_{top}|e_{spe})$ is the conditional probability that an SU pair can topologically connect to each other under the condition that the SU pair has spectrum available. Therefore, we need to derive the analytical expressions of $p(e_{spe})$ and $p(e_{top}|e_{spe})$ in order to calculate $p_{con}$. In particular, we analyse $p(e_{spe})$ in Section \ref{subsec:spectr_avail}, then derive $p(e_{top}|e_{spe})$ in Section \ref{subsec:top_conn}. Finally, we obtain closed-form expressions of $p_{con}$ in Section \ref{subsec:con_prob}.

\subsection{Spectrum Availability}
\label{subsec:spectr_avail}

Since there is no overlapping in the communication regions of any two SUs in a sparse network, the analysis on an SU pair applies to any SU pair. We denote the probability that an SU pair $(i,j)$ has spectrum available by $p_{ij}$. It follows that $p(e_{spe})=p_{ij}$.

\begin{figure}[t]
\centering
\subfigure[In Dir-CRNs and Omn-Dir-CRNs]{
\begin{minipage}{4.4cm}
\centering
\includegraphics[width=3.8cm]{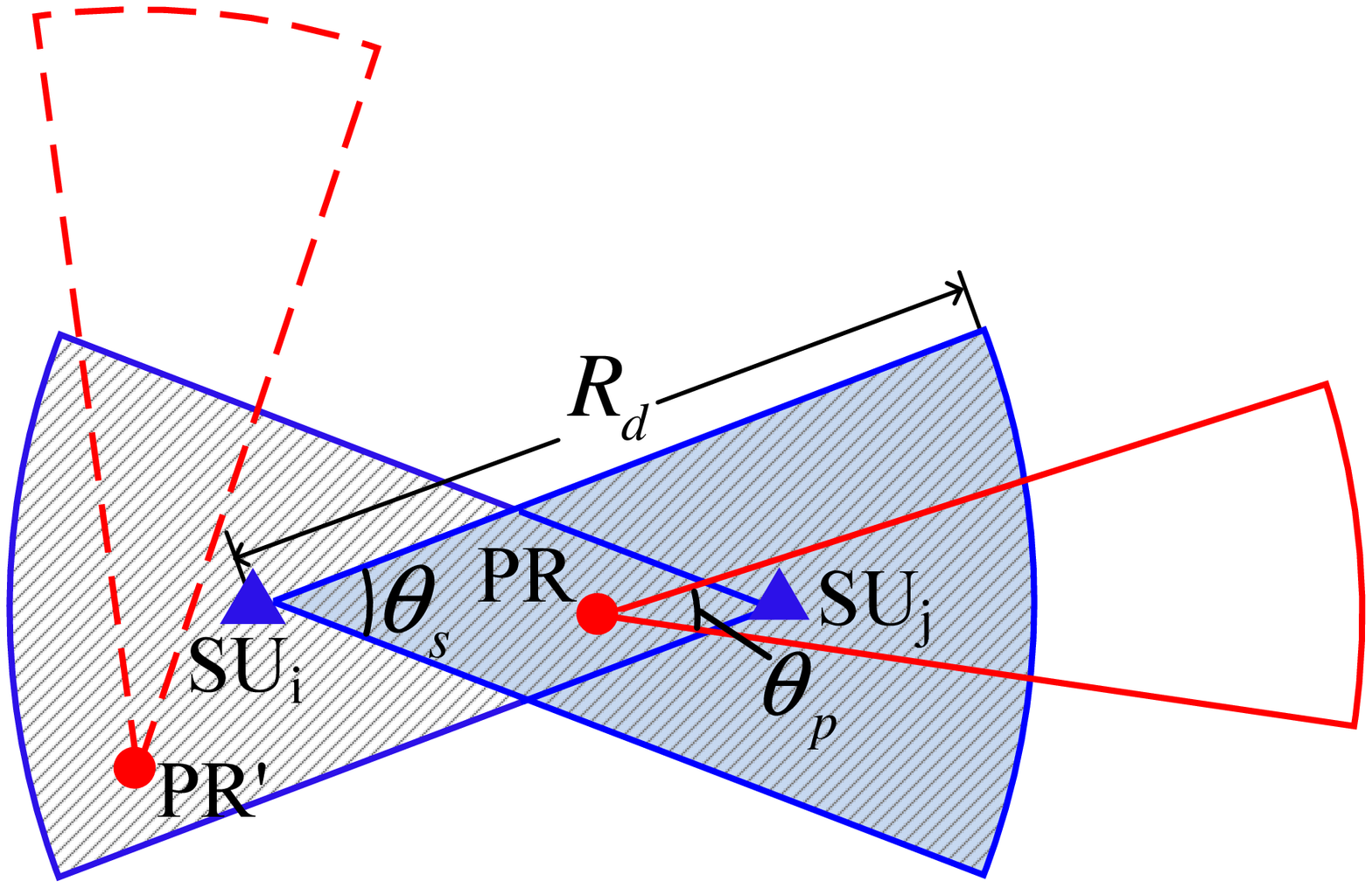}
\end{minipage}
\label{fig:dir}
}
\subfigure[In Omn-CRNs]{
\begin{minipage}{3.6cm}
\includegraphics[width=2.8cm]{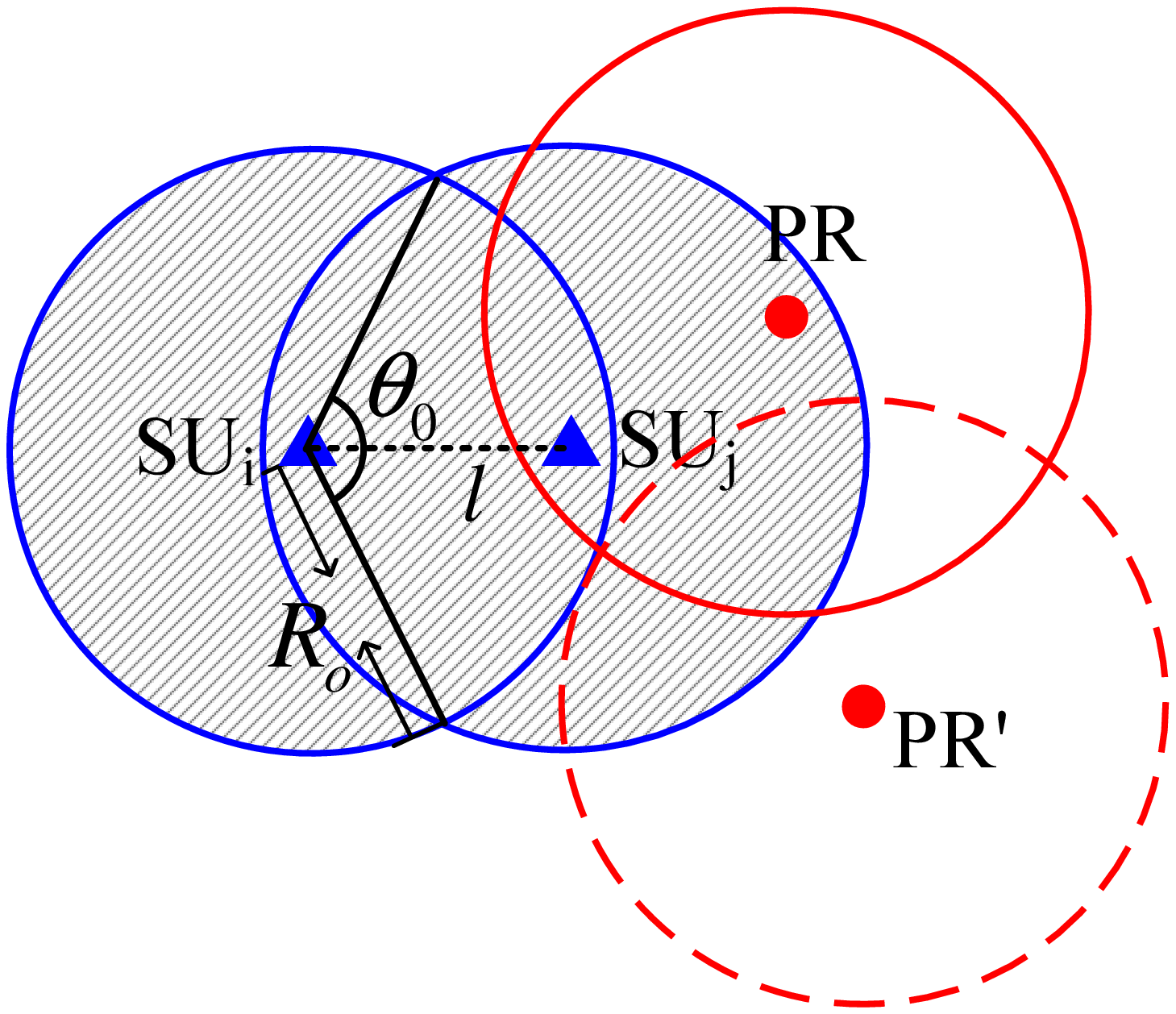}
\end{minipage}
\label{fig:omn}
}
\caption{Detection region of an SU pair}
\label{fig:supair}
\vspace*{-0.5cm}
\end{figure}

In order to derive $p_{ij}$, we first give the condition of a pair of SUs having spectrum available. According to the detect-and-avoid protocol as described in Section \ref{sec:int_contraint}, an SU pair (SU$_i$ and SU$_j$) has spectrum available if both the following conditions are satisfied:
\begin{enumerate}
\item No PR in the detection region of SU$_i$ covers SU$_i$;
\item No PR in the detection region of SU$_j$ covers SU$_j$.
\end{enumerate}
The \emph{detection region} is a region in which SUs cannot receive the detection preamble from PRs. Fig. \ref{fig:supair} shows an example of the detection region of an SU pair (SU$_i$ and SU$_j$). In particular, Fig. \ref{fig:dir} and Fig. \ref{fig:omn} show two different cases of Dir-CRNs (or Omn-Dir-CRNs) and Omn-CRNs, respectively. Take Fig. \ref{fig:dir} as an example, in which the existence of PR$'$ cannot affect SUs because the antenna beamwidth of PR$'$ does not cover SU$_i$ even if PR$'$ falls into the detection region of SU$_i$. However, there is a different case in Fig. \ref{fig:omn}, which requires no PRs falling into the detection region of SU$_i$ or the detection region of SU$_j$. Regarding to Omn-Dir-CRNs, we can have the similar case while we only need to consider the antenna direction of SUs since PUs transmit omni-directionally.

As shown in Fig. \ref{fig:dir}, the \emph{detection region} in Dir-CRNs is the union of two sectors (i.e., the shaded area in Fig. \ref{fig:dir}), each of which is bounded by the \emph{detection range} $R_d$ and the beamwidth $\theta_s$.  The detection range $R_d$ is the \emph{maximum detection distance} between an SU and a PR, which can be obtained by letting the left-hand-side (LHS) of Inequality (\ref{eq:SINR_d}) be equal to the right-hand-side (RHS) and replacing the corresponding antenna gains in Inequality (\ref{eq:SINR_d}) by $G_p$, $G_s$ and $G_o=1$, respectively.

Different from Dir-CRNs, the detection region of Omn-CRNs is the union of two circles (i.e., the shaded area in Fig. \ref{fig:omn}), each of which is bounded by the detection range denoted by $R_o$, which can be obtained by a similar approach to $R_d$. We then have $R_d$ and $R_o$ as follows, respectively,
\begin{equation}
\small
\begin{aligned}
R_d=\left(\frac{4\pi^2P_{d}h}{\theta_p\theta_s\eta}\right)^\frac{1}{\alpha}; R_o=\left(\frac{P_{d}h}{\eta}\right)^\frac{1}{\alpha}.
\label{eq:rd}
\end{aligned}
\end{equation}

We then analyse $p_{ij}$ according to different cases of Dir-CRNs, Omn-CRNs and Omn-Dir-CRNs. We denote the probability that an SU pair has spectrum available in Dir-CRNs by $p_{ij}^{d}$ and the probability that an SU pair has spectrum available in Omn-CRNs by $p_{ij}^{o}$. We first obtain the following result for $p_{ij}^{d}$.
\begin{theorem}
The probability that an SU pair has spectrum available in Dir-CRNs $p_{ij}^{d}$ is given as follows,
\begin{equation}
\small
p_{ij}^{d}=\exp\left(-\frac{\lambda_p}{2\pi}\left(\frac{4\pi^2 P_{d}}{\eta}\right)^\frac{2}{\alpha}(\theta_p\theta_s)^{\left(1-\frac{2}{\alpha}\right)}\Gamma\left(1+\frac{2}{\alpha}\right)\right)
\label{eq:pijd},
\end{equation}
where $\Gamma(\cdot)$ is the gamma function.
\label{theo:pij}
\end{theorem}

\textit{Proof:}
Since PRs are randomly distributed according to homogeneous PPP in density $\lambda_p$ and the antenna direction of each PR is uniformly distributed within $[0, 2\pi]$, the event that PRs can cover the SU$_i$ or SU$_j$ follows a homogeneous PPP with density $\lambda_p\cdot\frac{\theta_p}{2\pi}$. Therefore, $p_{ij}^d$ can be expressed as the following equation,
\begin{equation}
\small
p_{ij}^{d}=\left(\frac{(\lambda_p\cdot\frac{\theta_p}{2\pi}\cdot \mathbb{E}[S_d])^0}{0!}e^{-\lambda_p\cdot\frac{\theta_p}{2\pi}\cdot \mathbb{E}[S_d]}\right)^2=\exp\left(-\frac{\lambda_p\theta_p \mathbb{E}[S_d]}{\pi}\right),
\label{eq:pijd2}
\end{equation}
where $\mathbb{E}[S_d]$ is the expected value of the area of detection region of an SU in Dir-CRNs (e.g., the shaded region shown in Fig. \ref{fig:dir}), which is given by
\begin{equation}
\small
\mathbb{E}[S_{d}]=\mathbb{E}\left[\pi R_d^2\cdot\frac{\theta_s}{2\pi}\right]=\frac{\theta_s}{2}\mathbb{E}[R_d^2].
\label{eq:Sd}
\end{equation}

Substituting Eq. (\ref{eq:rd}) into Eq. (\ref{eq:Sd}), $\mathbb{E}[S_d]$ can be expressed as follows,
\begin{equation}
\small
\begin{aligned}
\mathbb{E}[S_{d}]&=\frac{\theta_s}{2}\left(\frac{4\pi^2P_{d}}{\theta_s\theta_p\eta}\right)^\frac{2}{\alpha}\int_{0}^{\infty}h^\frac{2}{\alpha}e^{-h}dh=\frac{\theta_s}{2}\left(\frac{4\pi^2P_{d}}{\theta_s\theta_p\eta}\right)^\frac{2}{\alpha}\Gamma\left(1+\frac{2}{\alpha}\right).
\label{eq:Esd}
\end{aligned}
\end{equation}

Combining Eq. (\ref{eq:Esd}) with Eq. (\ref{eq:pijd2}), we can obtain the results of Eq. (\ref{eq:pijd}).
\done

We can extend Theorem \ref{theo:pij} to the case of Omn-Dir-CRNs by setting $\theta_p=2\pi$ in Eq. (\ref{eq:pijd}). In particular, we have the following result.
\begin{corollary}
The probability that an SU pair has spectrum available in Omn-Dir-CRNs $p_{ij}^{od}$ is given as follows,
\begin{equation}
\small
p_{ij}^{od}=\exp\left(-\frac{\lambda_p}{2\pi}\left(\frac{4\pi^2 P_{d}}{\eta}\right)^\frac{2}{\alpha}(2\pi \theta_s)^{\left(1-\frac{2}{\alpha}\right)}\Gamma\left(1+\frac{2}{\alpha}\right)\right)
\label{eq:pijod}.
\end{equation}
\end{corollary}

\begin{remark}
Theorem \ref{theo:pij} shows that $p_{ij}^{d}$ depends on multiple factors such as $\lambda_p$, $\eta$, $\theta_p$, $\theta_s$ and $\alpha$. In particular, $p_{ij}^d$ is decreasing when $\lambda_p$ increases. In addition, when $\alpha=2$, $p_{ij}^{d}$ is independent of $\theta_p$ and $\theta_s$ since $(1-\frac{2}{\alpha})$ becomes 0. However, when $\alpha>2$, $p_{ij}^{d}$ decreases with increased factor $\theta_p\theta_s$, implying that narrower antenna beamwidth of PUs and SUs brings higher spectrum availability of any SU pair because of fewer SUs that receive detection preambles from PRs. Our numerical results will further confirm this observation.
\end{remark}

We then have the result on $p_{ij}^{o}$ as the following theorem.
\begin{theorem}
The probability that an SU pair can have the spectrum in Omn-CRNs $p_{ij}^{o}$ is given as follows,
\begin{equation}
\small
p_{ij}^{o}=\exp\left(-\left(\pi+\frac{3\sqrt{3}}{4}\right)\left(\frac{P_d}{\eta}\right)^{\frac{2}{\alpha}}\lambda_p\Gamma\left(1+\frac{2}{\alpha}\right)\right).
\label{eq:pijo}
\end{equation}
\end{theorem}

\textit{Proof:}
Since PRs in Omn-CRNs are equipped with omni-directional antennas, which can cover all directions, the PRs falling in the detection region of SU$_i$ or SU$_j$ definitely cover SU$_i$ or SU$_j$. Therefore, if both SU$_i$ and SU$_j$ have spectrum available, there must be no PRs in their detection regions. Then, $p_{ij}^o$ can be expressed as the following equation,
\begin{equation}
\small
p_{ij}^{o}=\frac{(\lambda_p\cdot \mathbb{E}[S_o])^0}{0!}e^{-\lambda_p\cdot \mathbb{E}[S_o]}=\exp\left(-\lambda_p\mathbb{E}[S_o]\right),
\label{eq:pijo2}
\end{equation}
where $\mathbb{E}[S_o]$ is the expected value of the area of detection region of SU$_i$ union the area of the detection region of SU$_j$ in Omn-CRNs (i.e., the shaded region shown in Fig. \ref{fig:omn}).

Note that $S_o$ is a random variable depending on the distance between SU$_i$ and SU$_j$ denoted by $l$ (as shown in Fig. \ref{fig:omn}). Then $S_o$ can be expressed by

\begin{equation}
\small
S_o(l)=(2\pi-\theta_0)R_o^2+lR_o\sin\frac{\theta_0}{2},
\label{eq:So}
\end{equation}
where $\theta_0=2\arccos\frac{l}{2R_o}$.

For simplification, we assume that the maximum value of $l$ is equal to the detection range $R_o$. According to the fact that each SR is uniformly randomly falling in the transmission region of a ST mentioned in Section \ref{Network Models}, the probability density function (PDF) of the distance $l$ can be expressed as
\begin{equation}
\small
f_l(l)=\frac{2\pi l}{\pi R_o^2}=\frac{2l}{R_o^2} \ (l\leq R_o).
\label{eq:fl}
\end{equation}

Then, $\mathbb{E}[S_o]$ can be expressed as
\begin{equation}
\footnotesize
\begin{aligned}
&\mathbb{E}[S_o]=\mathbb{E}_{h}(\mathbb{E}_l[S_o])=\mathbb{E}_{h}\left(\int_{0}^{R_o} {S_o(l)} \cdot {f_l(l)}dl\right)\\
&=\int_{0}^\infty\left(\pi+\frac{3\sqrt{3}}{4}\right)R_o^2\cdot e^{-h}dh=\left(\pi+\frac{3\sqrt{3}}{4}\right)\left(\frac{P_d}{\eta}\right)^\frac{2}{\alpha}\Gamma\left(1+\frac{2}{\alpha}\right).
\label{eq:Eso}
\end{aligned}
\end{equation}
Combining Eq. (\ref{eq:Eso}) with Eq. (\ref{eq:pijo2}), we obtain the result.
\done


\begin{figure}[t]
\centering
\subfigure[$\alpha=3$]{
\begin{minipage}{4.1cm}
\centering
\includegraphics[width=4.1cm]{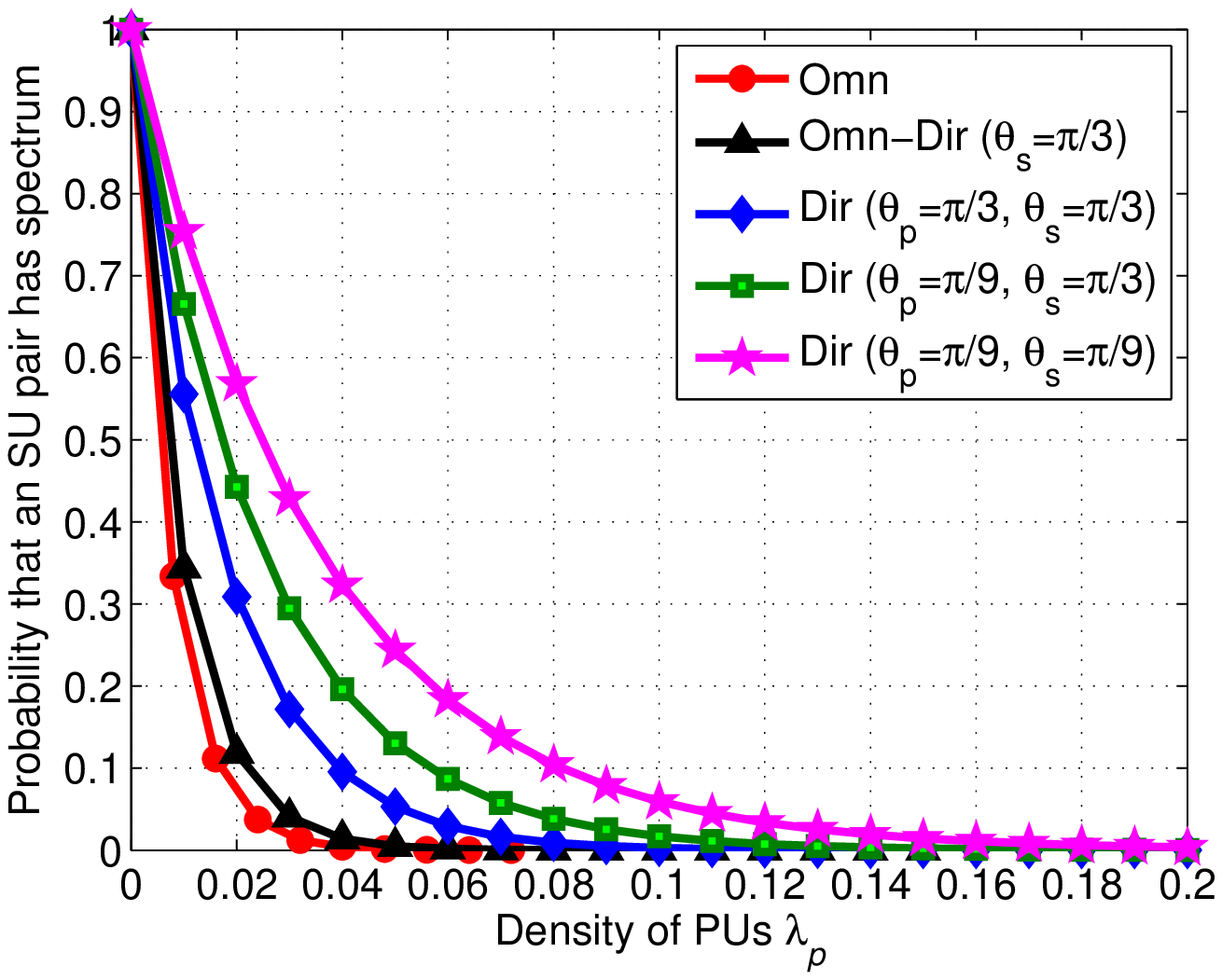}
\end{minipage}
\label{fig:pija3}
}\hfil
\subfigure[$\alpha=5$]{
\begin{minipage}{4.1cm}
\includegraphics[width=4.1cm]{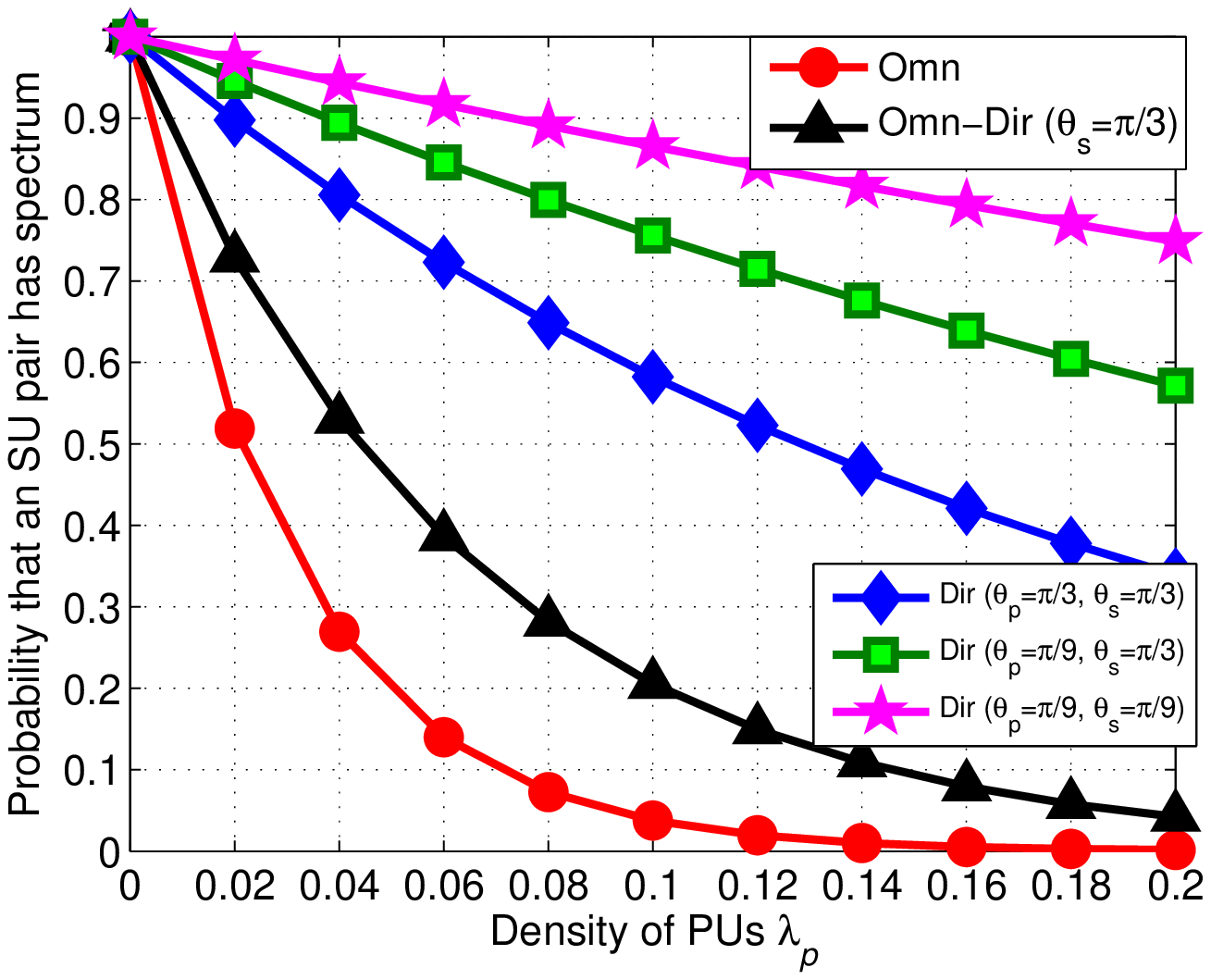}
\end{minipage}
\label{fig:pija5}
}\hfil
\caption{Probability that any SU pair has spectrum available versus $\lambda_p$ in Omn-CRNs, Dir-CRNs and Omn-Dir-CRNs (when $\theta_p=2\pi$) with different $F(\theta_s,\theta_p)$, where $P_d=10$.}
\label{fig:pij}
\vspace*{-0.5cm}
\end{figure}

Fig. \ref{fig:pij} plots the analytical results of $p_{ij}^{d}$, $p_{ij}^{od}$ and $p_{ij}^{o}$ with varied density of PUs $\lambda_p$ under different values of path loss exponent $\alpha$, in which Omn, Omn-Dir and Dir represent the results of Omn-CRNs, Omn-Dir-CRNs and Dir-CRNs, respectively. To investigate the impacts of antenna beamwidth on $p_{ij}^{d}$, we define $F(\theta_s,\theta_p)=\theta_s \cdot \theta_p$. Specifically, it is shown in Fig. \ref{fig:pij} that $p_{ij}$ decreases with the increased value of $\lambda_p$, implying that $p_{ij}$ heavily depends on the density of active PUs. In addition, Fig. \ref{fig:pij} also indicates that the higher path loss $\alpha$ results in the higher $p_{ij}$. This is because that the higher path loss results in the smaller detection region. Moreover, in Dir-CRNs, $p_{ij}^{d}$ decreases when $F(\theta_s,\theta_p)$ decreases from $\frac{\pi^2}{27}$ (green curves) to $\frac{\pi^2}{9}$ (blue curves), but still higher than $p_{ij}^{od}$ and $p_{ij}^{o}$. This indicates that Dir-CRNs can obtain a higher spectrum availability than Omn-Dir-CRNs and Omn-CRNs. More specifically,  Eq. (\ref{eq:pijd}) indicates that $p^d_{ij}$ is an increasing function of factor $\theta_s \cdot \theta_p$ (i.e., $F(\theta_s,\theta_p)$) when $\alpha>2$ (which is quite common in a realistic environment). This observation implies that decreasing the beamwidths of PUs and SUs can improve the spectrum availability, consequently contributing to the connectivity improvement of Dir-CRNs.


\subsection{Topological Connectivity}
\label{subsec:top_conn}

We next analyse the conditional probability $p(e_{top}|e_{spe})$. To simplify the derivation, we use $p_{top}$ to represent $p(e_{top}|e_{spe})$. We assume that each ST has the same transmission power $P_s$. The probability of topological connection $p_{top}$ of any SU pair can be expressed as follows,
\begin{equation}
\footnotesize
p_{top}=\mathbb{P}[{\text{SINR}} \geqslant \delta]=\mathbb{P}\left[\frac{P_{s}r^{-\alpha }hG_s^2}{I_{s}+I_{p}+\sigma ^{2}}\geqslant \delta\right],
\end{equation}
where SINR is the signal-to-interference-plus-noise ratio at a reference SU (named as SR$_0$), $\delta$ is the threshold that a link can be established, $\sigma^2$ is the thermal noise, $I_s$ and $I_p$ denote the cumulative interference generated from active STs to SR$_0$ and the cumulative interference from PTs to SR$_0$, respectively.

In particular, $I_p$ can be expressed as follows,
\begin{equation}
\footnotesize
I_{p}=\sum_{k=1}^n P_{p}{r_{k}^\prime}^{-\alpha}h_{k}^\prime G_{s_k}^\prime \widetilde G_{p_k}^{\prime},
\end{equation}
where $n$ is the number of PTs, $P_p$ is the transmitting power of each PT, $r_k^\prime$ is the distance from PT$_k$ to SR$_0$, $h_k^\prime$ is the Rayleigh fading factor in the channel between PT$_k$ and SR$_j$, $G_{s_k}^\prime$ is the antenna gain of SR$_0$ in the direction of PT$_k$, and $\widetilde G_{p_k}^{\prime}$ is the antenna gain of PT$_k$ in the direction of SR$_0$.

Moreover, $I_s$ can be expressed as follows,
\begin{equation}
\footnotesize
I_{s}=\sum_{g=1}^mP_{s}{r_{g}}^{-\alpha}h_{g}G_{s_g}\widetilde G_{s_g},
\end{equation}
where $m$ is the number of active STs, $P_s$ is the transmitting power of each ST, $r_g$ is the distance from ST$_g$ to SR$_0$, $h_g$ is the Rayleigh fading factor in the channel between PT$_g$ and SR$_0$, $G_{s_g}$ is the antenna gain of SR$_0$ in the direction of ST$_g$, and $\widetilde G_{s_g}$ is the antenna gain of ST$_g$ in the direction of SR$_0$.

In order to derive both $I_{p}$ and $I_{s}$, we need to investigate both the distribution of PTs and the distribution of active STs. In particular, the distribution of PTs is affected by the condition that an SU pair has spectrum available. More specifically, Section \ref{subsec:spectr_avail} shows that an SU pair has spectrum available implying no PRs in the detection region of the SU pair. This condition not only restricts the locations of PRs but also confines the locations of PTs due to the correlation between a PT and a PR. Therefore, we can have the approximation that PTs also follow a homogeneous PPP in the whole network under the condition that the SU pair has spectrum available since the detection region of an SU pair is too small compared with the whole network.

With respect to the distribution of STs in Dir-CRNs and Omn-CRNs, we have the following results.
\begin{lemma}
\label{lemma:dist_st}
The active STs of Dir-CRNs follows a thinning homogeneous PPP with density $\lambda _{s}^d$, which can be expressed as the following equation,
\begin{equation}
\footnotesize
\begin{aligned}
\lambda _{s}^d&=\lambda_s\cdot p_{ij}^{d}=\lambda_s\cdot\exp\left(-\frac{\lambda_p}{2\pi}\left(\frac{4\pi^2 P_{d}}{\eta}\right)^\frac{2}{\alpha}(\theta_p\theta_s)^{\left(1-\frac{2}{\alpha}\right)}\Gamma\left(1+\frac{2}{\alpha}\right)\right).
\end{aligned}
\end{equation}
The active STs of Omn-CRNs also follows a thinning homogeneous PPP with density $\lambda _{s}^o$, which can be expressed as the following equation,
\begin{equation}
\footnotesize
\begin{aligned}
\lambda _{s}^o&=\lambda_s\cdot p_{ij}^{o}=\lambda_s\cdot\exp\left(-\left(\pi+\frac{3\sqrt{3}}{4}\right)\left(\frac{P_d}{\eta}\right)^{\frac{2}{\alpha}}\lambda_p\Gamma\left(1+\frac{2}{\alpha}\right)\right).
\end{aligned}
\end{equation}
\end{lemma}

\begin{proof}
The detailed proof is presented in \cite{DirCRN:2018}.
\end{proof}

We then have the probability of topological connection of any SU pair, denoted by $p_{top}^d$ as follows.
\begin{theorem}
The probability of topological connection of any SU pair in Dir-CRNs is given by
\begin{equation}
\footnotesize
p_{top}^d=\exp \left(-\frac{\delta\sigma ^{2}r^{\alpha }\theta_s^2}{4\pi^2P_{s}}-\frac{\delta ^{%
\frac{2}{\alpha }}r^{2}(\lambda _{p}\theta _{s}^{1+\frac{2}{\alpha
}}\theta _{p}^{1-\frac{2}{\alpha }}(\frac{P_p}{P_s})^\frac{2}{\alpha}+\lambda _{s}p_{ij}^{d}\theta _{s}^{2})}{%
2\alpha \sin (\frac{2\pi }{\alpha })}\right),
\label{eq:ptopd}
\end{equation}
where $ p_{ij}^{d}=\exp\left(-\frac{\lambda_p}{2\pi}\left(\frac{4\pi^2 P_{d}}{\eta}\right)^\frac{2}{\alpha}(\theta_p\theta_s)^{\left(1-\frac{2}{\alpha}\right)}\Gamma\left(1+\frac{2}{\alpha}\right)\right)$ as given by Eq. (\ref{eq:pijd}).
\label{theo:pijd}
\end{theorem}

\begin{proof}
The probability of topological connection of any SU pair in Dir-CRNs is
\begin{equation}
\footnotesize
\begin{aligned}
&p_{top}^d=\mathbb{P}\left[\frac{P_{s}r^{-\alpha }hG_{s}^2}{I_{s}+I_{p}+\sigma ^{2}}%
\geqslant \delta \right]=\mathbb{P}\left[h\geqslant \frac{\delta (I_{s}+I_{p}+\sigma ^{2})}{%
G_{s}G_{s}P_{s}r^{-\alpha }}\right] \\
&=\exp\left(-\frac{\delta \sigma ^{2}}{G_{s}^2P_{s}r^{-\alpha }}\right)\mathcal{L}_{I_{p}}\left(%
\frac{b}{P_{s}}\right)\mathcal{L}_{I_{s}}\left(\frac{b}{%
P_{s}}\right),
\label{eq:topconnectivity}
\end{aligned}
\end{equation}
where $b=\frac{\delta }{G_{s}^2r^{-\alpha }}$ and $\mathcal{L}_{X}(x)$ is the Laplace transform of random variable $X$ at $x$.

The Laplace transform $\mathcal{L}_{I_{p}}(\frac{b}{P_{s}})$ in Eq. (\ref{eq:topconnectivity}) can be calculated as follows,
\begin{equation}
\footnotesize
\begin{aligned}
&\mathcal{L}_{I_{p}}\left(\frac{b}{P_{s}}\right)=\mathbb{E}_{I_p}[e^{-\frac{bI_p}{P_s}}]=\mathbb{E}_{h_k^{\prime }, r_{k}^{\prime }, G_{s_k}^{\prime }\widetilde G_{p_k}^{\prime }}[e^{-\frac{bP_p}{P_s}\sum_{k=1}^n
{r_{k}^{\prime }}^{-\alpha }h_{k}^{\prime }G_{s_k}^{\prime }\widetilde G_{p_k}^{\prime }}] \\
&=\mathbb{E}_{r_{k}^{\prime },G_{sk}^{\prime }\widetilde G_{p_k}^{\prime }}\prod_{k=1}^n \mathbb{E}_{h}[e^{-\frac{bP_p}{P_s}{r_{k}^{\prime }}^{-\alpha
}hG_{s_k}^{\prime }\widetilde G_{p_k}^{\prime }}],
\label{eq:LIp1}
\end{aligned}
\end{equation}
where the last step is obtained from the fact that random variables $h_k^{\prime }$, $r_k^{\prime }$ and $G_{s_k}^{\prime }G_{p_k}^{\prime }$ are mutually independent and Rayleigh fading variable $h_k^{\prime }$ is i.i.d. for $1\leq k \leq n$.

The probability generation function (PGF) of a PPP (denoted by $\Phi$) in a space $S$ has a property expressed as follows: for a function $0<f(x)<1$ ($x\in\Phi$), $\mathbb{E}[\prod_{x\in\Phi}f(x)]=\exp\left(-\lambda\int_S(1-f(x))dx\right)$ \cite{Kingman:1993}. We then use this property and extend it to $R^2\times[0,2\pi]$ with density $\lambda_p/2\pi$ (with consideration of antenna orientation $\varphi_k$) \cite{Georgiou:2015}. We next represent Eq. (\ref{eq:LIp1}) as follows,
\begin{equation}
\footnotesize
\mathcal{L}_{I_{p}}\left(\frac{b}{P_{s}}\right)=\exp \left(-\frac{\lambda _{p}}{2\pi }\int_0^{2\pi} \int_{R^2}
1-\mathbb{E}_{h}[e^{-\frac{bP_p}{P_s}{r_{k}^{\prime }}^{-\alpha }hG_{s_k}^{\prime }\widetilde G_{p_k}^{\prime
}}]d r_{k}^{\prime }d\varphi_{k}\right),
\label{eq:LIp2}
\end{equation}
where $r_k^{\prime}$ is the relative position of PT$_k$ in the direction of SR$_0$ and the expectation $\mathbb{E}_h[e^{-\frac{bP_p}{P_s}{r_{k}^{\prime}}^{-\alpha}hG_{s_k}^{\prime }\widetilde G_{p_k}^{\prime}}]$ can be calculated as follows,
\begin{equation*}
\footnotesize
\begin{aligned}
\mathbb{E}_h[e^{-\frac{bP_p}{P_s}{r_{k}^{\prime}}^{-\alpha}hG_{s_k}^{\prime }\widetilde G_{p_k}^{\prime}}]&=\int_0^{\infty}e^{-\frac{bP_p}{P_s}{r_k^{\prime }}^{-\alpha}hG_{s_k}^{\prime }\widetilde G_{p_k}^{\prime }}\cdot e^{-h}dh\\
&=\frac{1}{\frac{bP_p}{P_s}{r_{k}^{\prime }}^{-\alpha }G_{s_k}^{\prime }\widetilde G_{p_k}^{\prime }+1}.
\end{aligned}
\end{equation*}

\begin{figure}[t]
\centering
\includegraphics[width=6cm]{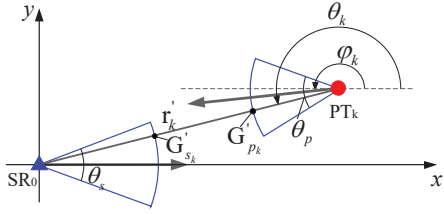}
\caption{Schematic illustration of SR$_0$ and PT$_k$}
\label{fig:srpt}
\vspace*{-0.5cm}
\end{figure}

Fig. \ref{fig:srpt} gives us a schematic illustration of SR$_0$ and PT$_k$ in a coordinate system, in which both antenna gains $G_{s_k}^{\prime }$ and $G_{p_k}^{\prime }$ are determined by $\varphi_k$ and $\theta_k$, where $\theta_k$ denotes the angle from $x$-axis to the relative direction of SR$_0$ in the direction of PT$_k$. Therefore, Eq. (\ref{eq:LIp2}) can be expressed as follows,
\begin{equation}
\footnotesize
\begin{aligned}
&\mathcal{L}_{I_{p}}\left(\frac{b}{P_{s}}\right)=\exp \left(-\frac{\lambda _{p}}{2\pi }\int_0^{2\pi} \int_{R^2}
\frac{\frac{bP_p}{P_s}{r_{k}^{\prime }}^{-\alpha }G_{s_k}^{\prime }\widetilde G_{p_k}^{\prime }}{\frac{bP_p}{P_s}{r_{k}^{\prime }}^{-\alpha }G_{s_k}^{\prime }\widetilde G_{p_k}^{\prime }+1}%
d\mathbf r_{k}^{\prime }d\varphi _{k}^{{}}\right) \\
&=\exp \left(-\frac{\lambda _{p}}{\pi }\int_0^\infty\int_0^\frac{\theta_s}{2} \int_{\theta_k-\frac{\theta_p}{2}}^{\theta_k+\frac{\theta_p}{2}}
\frac{\frac{bP_p}{P_s}{r_{k}^{\prime }}^{1-\alpha }G_{s}G_{p}}{\frac{bP_p}{P_s}{r_{k}^{\prime }}^{-\alpha }G_{s}G_{p}+1}%
d\varphi_{k}d\theta_kdr_{k}^{\prime }\right) \\
&=\exp \left(-\frac{\lambda _{p}(2\pi )^{\frac{4}{\alpha }}\left(\frac{bP_p}{P_s}\right)^{\frac{2}{\alpha }%
}(\theta _{s}\theta _{p})^{1-\frac{2}{\alpha }}}{2\alpha \sin \left(\frac{2\pi }{%
\alpha }\right)}\right).
\label{eq:LIp3}
\end{aligned}
\end{equation}

Following the similar derivation procedure, we have the $\mathcal{L}_{I_{s}}(\frac{b}{P_{s}})$ as follows,
\begin{equation}
\footnotesize
\begin{aligned}
&\mathcal{L}_{I_{s}}\left(\frac{b}{P_{s}}\right)=\mathbb{E}_{I_s}[e^{-\frac{bI_s}{P_s}}]=\mathbb{E}_{h_g,r_{g},G_{s_g}\widetilde G_{s_g}}[e^{-b\sum_{g=1}^m
r_{g}^{-\alpha }h_{g}G_{s_g}\widetilde G_{s_g}}] \\
&=\exp \left(-\frac{\lambda _{s}p_{ij}^s(2\pi )^{\frac{4}{\alpha }}b^{%
\frac{2}{\alpha }}(\theta _{s})^{2-\frac{4}{\alpha }}}{2\alpha \sin \left(\frac{%
2\pi }{\alpha }\right)}\right),
\end{aligned}
\label{eq:LIs}
\end{equation}
where $r_g$ is the relative position of ST$_g$ in the direction of SR$_0$, $\varphi_g$ is the antenna orientation of ST$_g$ and $\theta_g$ is the angle from $x$-axis to the relative direction of SR$_0$ in the direction of ST$_g$.

%


%

Combining Eq. (\ref{eq:LIs}), Eq. (\ref{eq:LIp3}) and Eq. (\ref{eq:topconnectivity}), we have the topological connectivity of any SU pair in Dir-CRNs given by Eq. (\ref{eq:ptopd}).
\end{proof}

\begin{table*}[t]
\begin{equation}
\footnotesize
\begin{aligned}
p_{con}^{d}=p_{top}^{d}\cdot p_{ij}^d=\exp\left(-\left(\frac{4\pi^2P_{d}}{\theta_p\theta_s\eta}\right)^\frac{2}{\alpha}\frac{\lambda_p\theta_p\theta_s}{2\pi}\Gamma\left(\frac{2+\alpha}{\alpha}\right)-\frac{\delta\sigma ^{2}r^{\alpha}\theta_s^2}{4\pi^2P_{s}}-\frac{\delta ^{
\frac{2}{\alpha }}r^{2}(\lambda _{p}\theta _{s}^{1+\frac{2}{\alpha
}}\theta _{p}^{1-\frac{2}{\alpha }}(\frac{P_p}{P_s})^\frac{2}{\alpha}+p_{ij}^d\lambda _{s}\theta _{s}^{2})}{
2\alpha \sin (\frac{2\pi }{\alpha })}\right).
\end{aligned}
\label{eq:p_cond}
\end{equation}
\vspace*{-0.5cm}
\end{table*}

\begin{table*}[t]
\begin{equation}
\footnotesize
\begin{aligned}
p_{con}^{od}=p_{top}^{od}\cdot p_{ij}^{od}=\exp\left(-\left(\frac{2\pi P_{d}}{\theta_s\eta}\right)^\frac{2}{\alpha}\lambda_p\theta_s\Gamma\left(\frac{2+\alpha}{\alpha}\right)-\frac{\delta\sigma ^{2}r^{\alpha}\theta_s^2}{4\pi^2P_{s}}-\frac{\delta ^{
\frac{2}{\alpha }}r^{2}(\lambda _{p}\theta _{s}^{1+\frac{2}{\alpha
}}(2\pi)^{1-\frac{2}{\alpha }}(\frac{P_p}{P_s})^\frac{2}{\alpha}+p_{ij}^{od}\lambda_{s}\theta _{s}^{2})}{
2\alpha \sin (\frac{2\pi }{\alpha })}\right).
\end{aligned}
\label{eq:p_condo}
\end{equation}
\vspace*{-0.5cm}
\end{table*}

\begin{table*}[t]
\begin{equation}
\footnotesize
\begin{aligned}
p_{con}^{o}=p_{top}^{o}\cdot p_{ij}^{o}=\exp\left(-\left(\pi+\frac{3\sqrt{3}}{4}\right)\left(\frac{P_d}{\eta}\right)^{\frac{2}{\alpha}}\lambda_p\Gamma\left(\frac{2+\alpha}{\alpha}\right)-\frac{\delta\sigma ^{2}r^{\alpha}}{P_{s}}-\frac{\delta ^{
\frac{2}{\alpha }}r^{2}\left(\lambda _{p}\left(\frac{P_p}{P_s}\right)^\frac{2}{\alpha}+\lambda _{s}p_{ij}^{o}\right)}{
\alpha \sin \left(\frac{2\pi }{\alpha }\right)}\right).
\end{aligned}
\label{eq:p_cono}
\end{equation}
\vspace*{-0.5cm}
\end{table*}

Essentially, Theorem \ref{theo:pijd} is a general expression of $p_{top}$, which can also be applied for Omn-Dir-CRNs and Omn-CRNs. In particular, we can obtain the probability of topological connection of any SU pair in Omn-Dir-CRNs denoted by $p_{top}^{od}$ by letting $\theta_p=2\pi$ and replacing $p_{ij}^d$ in in Eq. (\ref{eq:ptopd}) by $p_{ij}^{od}$ given in Eq. (\ref{eq:pijod}). This extension is given by the following corollary.
\begin{corollary}
The probability of topological connection of any SU pair in Omn-Dir-CRNs is given by
\begin{equation}
\footnotesize
p_{top}^{od}=\exp \left(-\frac{\delta\sigma ^{2}r^{\alpha }\theta_s^2}{4\pi^2P_{s}}-\frac{\delta ^{%
\frac{2}{\alpha }}r^{2}(\lambda _{p}\theta _{s}^{1+\frac{2}{\alpha
}}(2\pi)^{1-\frac{2}{\alpha }}(\frac{P_p}{P_s})^\frac{2}{\alpha}+\lambda _{s}p_{ij}^{od}\theta _{s}^{2})}{%
2\alpha \sin (\frac{2\pi }{\alpha })}\right),
\label{eq:ptopod}
\end{equation}
where $ p_{ij}^{od}$  is given by Eq. (\ref{eq:pijod}).
\label{cor:pijd}
\end{corollary}

Similarly, we denote the probability of topological connection of any SU pair in Omn-CRNs by $p_{top}^{o}$. After replacing $\theta_p=2\pi$, $\theta_s=2\pi$ and $p_{ij}^d$ in Eq. (\ref{eq:ptopd}) by $p_{ij}^o$ given in Eq. (\ref{eq:pijo}), we have the following corollary.
\begin{corollary}
The probability of topological connection of any SU pair in Omn-CRNs is
\begin{equation}
\footnotesize
p_{top}^{o}=\exp \left(-\frac{\delta\sigma ^{2}r^{\alpha}}{P_{s}}-\frac{\delta ^{
\frac{2}{\alpha }}r^{2}\left(\lambda _{p}\left(\frac{P_p}{P_s}\right)^\frac{2}{\alpha}+\lambda _{s}p_{ij}^{o}\right)}{
\alpha \sin \left(\frac{2\pi }{\alpha }\right)}\right),
\end{equation}
where $p_{ij}^{o}=\exp\left(-\left(\pi+\frac{3\sqrt{3}}{4}\right)\left(\frac{P_d}{\eta}\right)^{\frac{2}{\alpha}}\lambda_p\Gamma\left(1+\frac{2}{\alpha}\right)\right)$ as given by Eq. (\ref{eq:pijo}).
\end{corollary}

\begin{figure}[t]
\centering
\subfigure[$\alpha=3$]{
\begin{minipage}{4.1cm}
\centering
\includegraphics[width=4cm]{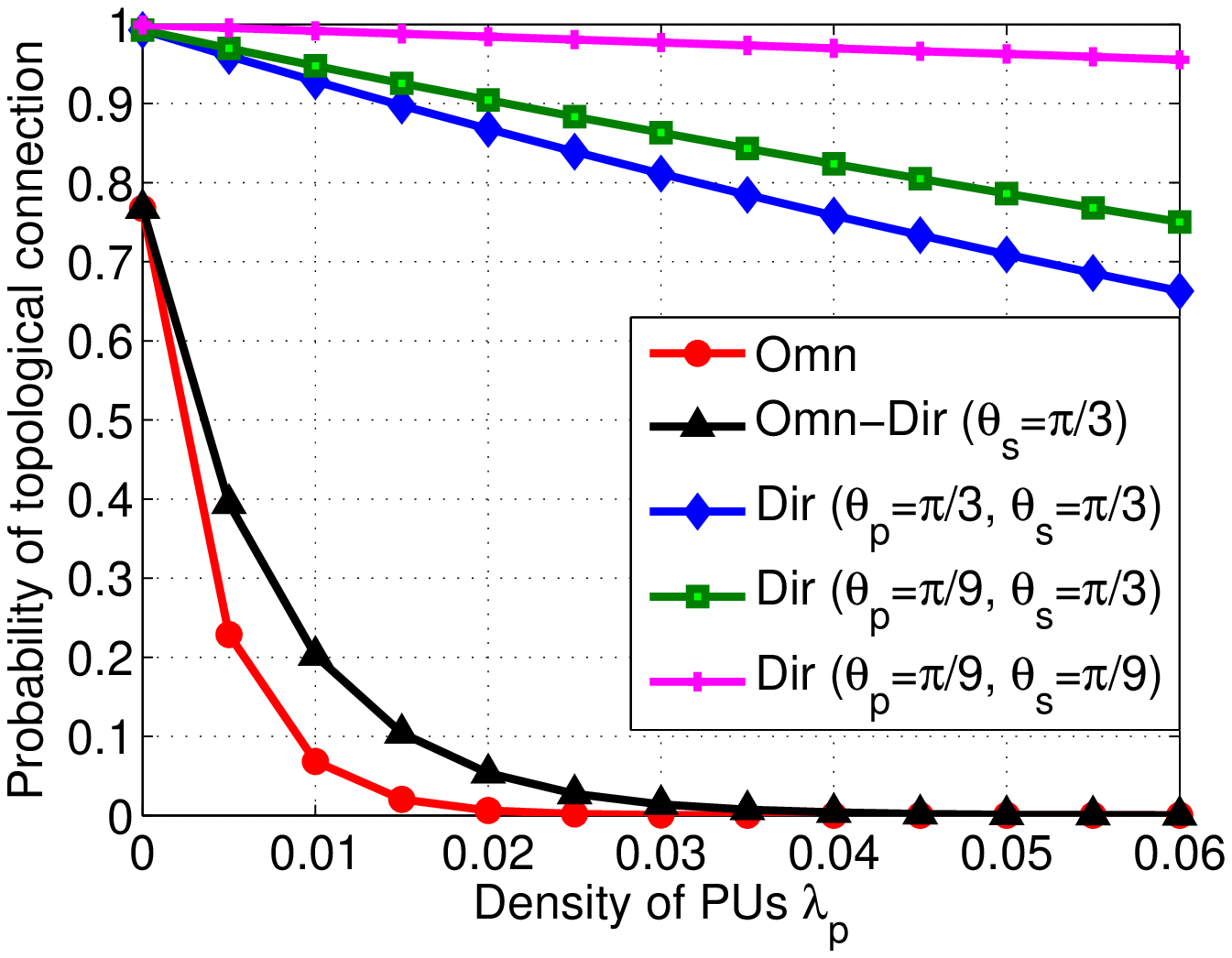}
\end{minipage}
\label{fig:ptopa3}
}\hfil
\subfigure[$\alpha=5$]{
\begin{minipage}{4.1cm}
\centering
\includegraphics[width=4cm]{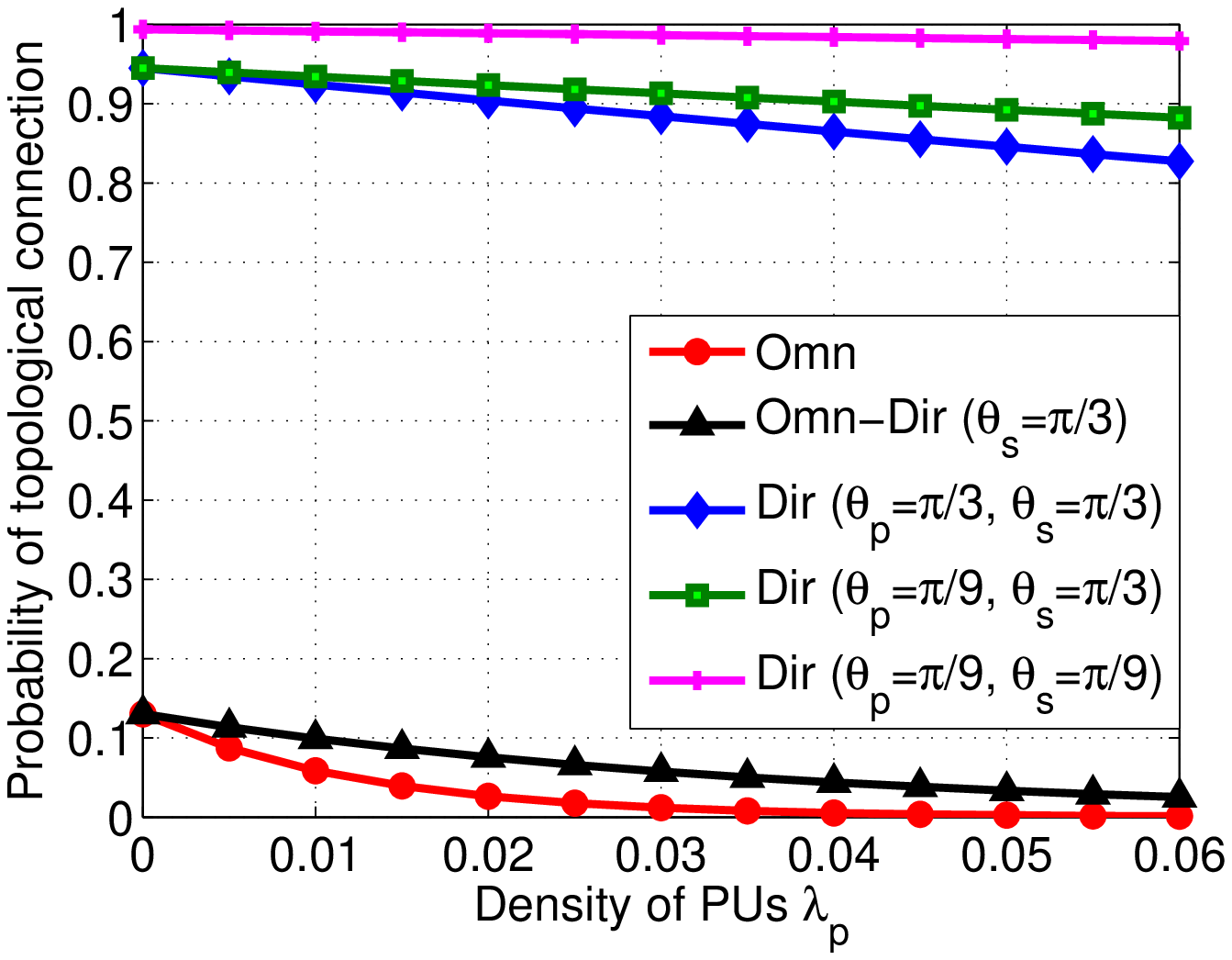}
 \end{minipage}
\label{fig:ptopa5}
}\hfil
\caption{Probability of topological connection of any SU pair versus the density of PUs $\lambda_p$ in Omn-CRNs, Omn-Dir-CRNs and Dir-CRNs, where $P_d=10$, $P_p=8$, $P_s=6$, $r=3$, $\lambda_s=0.0002$, $\eta=0.05$, $\sigma^2=0.01$, $\delta=5$.}
\label{fig:ptop}
\end{figure}


\begin{figure}[t]
\centering
\includegraphics[width=7.5cm]{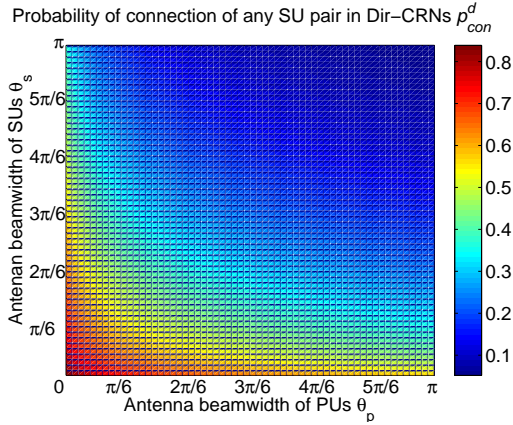}
\caption{Probability of connection of an SU pair versus in Dir-CRNs with different $\theta_p$ and $\theta_s$, where $P_d=10$, $P_p=8$, $P_s=6$, $\alpha=3$, $\lambda_p=0.02$, $\lambda_s=0.0002$, $\eta=0.05$, $\sigma^2=0.01$, $\delta=5$, $r=3$.}
\label{fig:beamwidth}
\vspace*{-0.5cm}
\end{figure}

Fig. \ref{fig:ptop} presents the analytical results of probability of topological connection of any SU pair versus distance $r$ with $\alpha=3$ and $\alpha=5$, respectively. In particular, Omn, Omn-Dir and Dir represent the results of Omn-CRNs, Omn-Dir-CRNs and Omn-CRNs, respectively. It is shown in Fig. \ref{fig:ptop} that $p_{top}^{d}$ of DIR-CRNs is significantly larger than that of Omn-Dir-CRNs and that of Omn-CRNs.

\begin{remark}
This result implies that \emph{using directional antennas in cognitive radio networks can significantly improve the probability of topological connection}. This improvement mainly owes to the higher SINR of Dir-CRNs than that of Omn-CRNs. Compared with omni-directional antennas, directional antennas can concentrate the transmission on desired directions while reducing the interference to other undesired directions. Meanwhile, \emph{using directional antennas at SUs can also improve the probability of topological connection}. This is because using directional antennas at SUs can extend the transmission range, consequently enhancing the probability of topological connection.
\end{remark}

We also compare the results with different values of $\theta_s$ and $\theta_p$ in Dir-CRNs. Fig. \ref{fig:ptop} shows that $p^d_{top}$ increases when $\theta_s$ is fixed at $\frac{\pi}{3}$ and $\theta_p$ decreases from $\frac{\pi}{3}$ to $\frac{\pi}{9}$ while $p^d_{top}$ increases when $\theta_p$ is fixed at $\frac{\pi}{9}$ and $\theta_s$ decreases from $\frac{\pi}{3}$ to $\frac{\pi}{9}$. This result indicates that \emph{using narrow-beamwidth antennas at both SUs and PUs can improve the probability of topological connection}. Compared with the case of using narrow-beamwidth antennas at PUs, using narrow-beamwidth antennas at SUs can further improve the probability of topological connection.

\subsection{Connectivity}
\label{subsec:con_prob}

Following the definition of the probability of connection of any SU pair $p_{con}=p(e_{spe}e_{top})=p(e_{spe})p(e_{top}|e_{spe})$ and the derivations of $p(e_{spe})$ and $p(e_{top}|e_{spe})$ in Section \ref{subsec:spectr_avail} and Section \ref{subsec:top_conn}, we then obtain the closed-form expressions of $p^d_{con}$ of Dir-CRNs, $p^{od}_{con}$ of Omn-Dir-CRNs and $p^o_{con}$ of Omn-CRNs as given in Eq. (\ref{eq:p_cond}), Eq. (\ref{eq:p_condo}) and Eq. (\ref{eq:p_cono}), respectively.

We further investigate the impacts of $\theta_p$ and $\theta_s$ on $p^d_{con}$ of Dir-CRNs. The effect of $\theta_p$ and $\theta_s$ on $p^d_{con}$ is illustrated in Fig. \ref{fig:beamwidth}. More specifically, it is shown in Fig. \ref{fig:beamwidth} that $p^d_{con}$ increases with the decreased value of $\theta_p$ and the decreased value of $\theta_s$, implying that using the narrow-beamwidth antennas at both SUs and PUs improves the connectivity of SUs. However, the narrower antenna beamwidth on the other hand can lead to the challenges in neighbor-discovery \cite{hndai:ijcs13} and blockage-effect reduction \cite{Bai:twireless15}. Therefore, there is a trade-off in choosing the beamwidth. How to choose the appropriate beamwidth at PUs and SUs is one of our future directions.

\section{Simulations}
\label{sec:sim}


\subsection{Simulation method}
\label{subsec:method}



We conduct extensive simulations to evaluate the accuracy of the proposed analytical models. We choose the sophisticated commercial software Matlab as the simulation tool. In particular, PUs are distributed according to HPPP in a plane of area $1,200 \times 1,200$. To eliminate the border effect, SUs are distributed according to HPPP on the sub-area with area $1,000 \times 1,000$ within the plane \cite{bettstetter:connectivity04,qwang:sensors2017}. The system parameters are chosen as $P_d=10$, $P_p=8$, $P_s=6$, $\eta=0.05$, $\sigma^2=0.01$, $\delta=5$, $\lambda_s=0.0002$. We denote the simulation result of the probability of connectivity by $p^s_{con}$ in order to differentiate it from the analytical result $p_{con}$. As shown in \cite{bettstetter:connectivity04,qwang:sensors2017}, $p^s_{con}$ is given by
\begin{equation}
\footnotesize
p^s_{con}=\frac{\textrm{\small \# topologies that an SU pair can connect successfully}}{\Omega},
\label{pcon_sim}
\end{equation}
where \# means the number of. In order to obtain an approximated result to the analytical one, we need to choose a large enough $\Omega$ (theoretically $\Omega\rightarrow\infty$) while it is extremely time-consuming to obtain the results of large $\Omega$. In this paper, we choose $\Omega=3,000$.

\subsection{Simulation results}
\label{subsec:pcon}

\begin{figure}[t]
\centering
\subfigure[$\alpha=3$]{
\begin{minipage}{4.1cm}
\includegraphics[width=4.1cm]{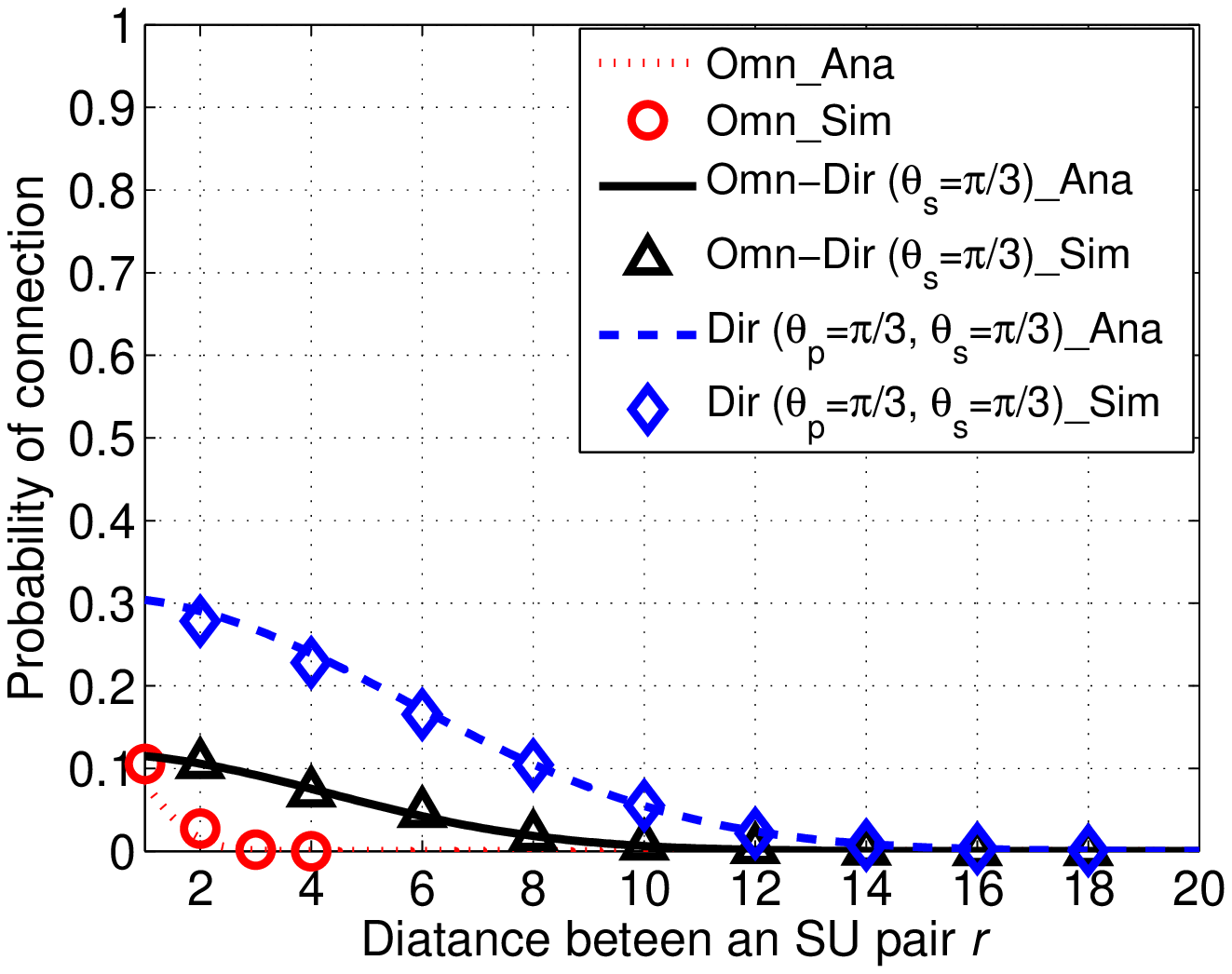}
\end{minipage}
\label{fig:pcon_r_a3}
}\hfil
\subfigure[$\alpha=5$]{
\begin{minipage}{4.1cm}
\includegraphics[width=4.1cm]{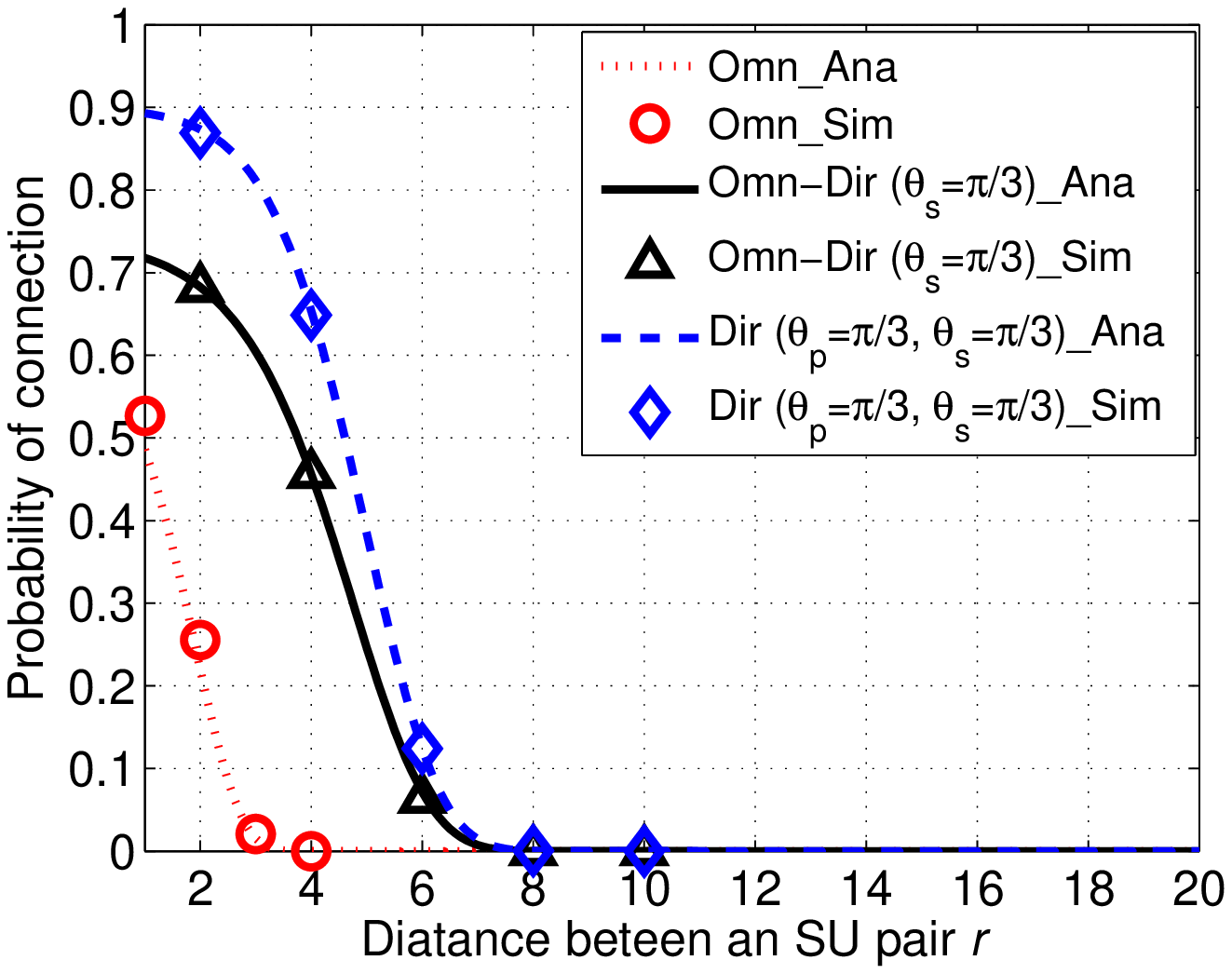}
\end{minipage}
\label{fig:pcon_r_a5}
}\hfil
\caption{Probability of connection of an SU pair versus the distance $r$ in Omn-CRNs, Omn-Dir-CRNs and Dir-CRNs, where $\lambda_p=0.02$.}
\label{fig:pcon}
\vspace*{-0.5cm}
\end{figure}

\begin{figure}[t]
\centering
\subfigure[$\alpha=3$]{
 \begin{minipage}{4.1cm}
\includegraphics[width=4.1cm]{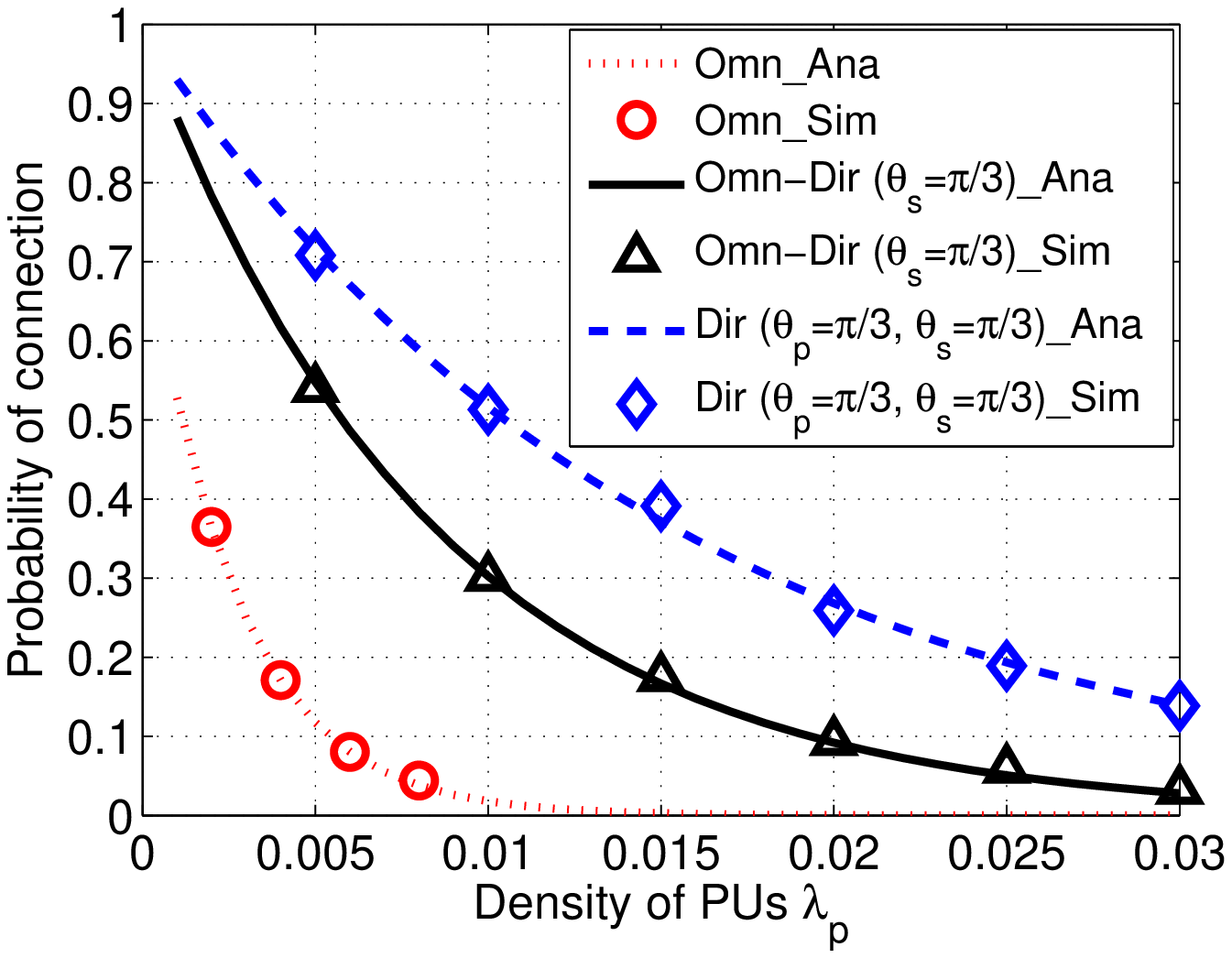}
\end{minipage}
\label{fig:pcona3_lambda_p}
}\hfil
\subfigure[$\alpha=5$]{
 \begin{minipage}{4.1cm}
\includegraphics[width=4.1cm]{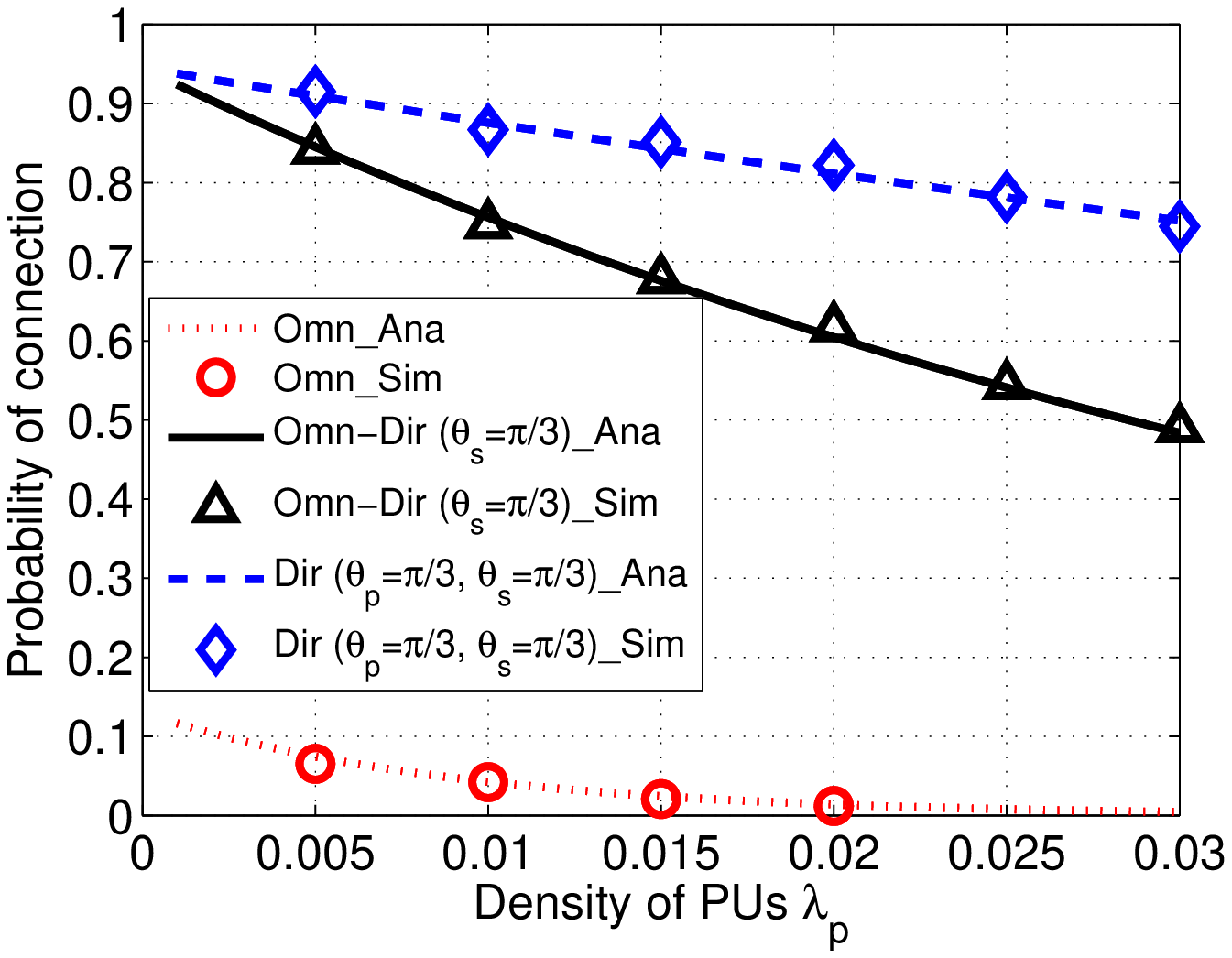}
\end{minipage}
\label{fig:pcona5_lambda_p}
}\hfil
\caption{Probability of connection of an SU pair versus the density of PUs $\lambda_p$ in Omn-CRNs, Omn-Dir-CRNs and Dir-CRNs, where $r=3$.}
\label{fig:pcon_lambda_p}
\vspace*{-0.5cm}
\end{figure}

\subsubsection{Impact of distance $r$}

In the first set of simulations, we investigate impact of the distance $r$ on the probability of connection of an SU pair in Omn-CRNs, Omn-Dir-CRNs and Dir-CRNs. Fig. \ref{fig:pcon} presents the results, in which the analytical results are plotted as curves and the simulation results are represented as markers. It is shown in Fig. \ref{fig:pcon} that there is an excellent agreement of the simulation results with the analytical results, implying that our proposed analytical model is fairly accurate.

Meanwhile, it is shown in Fig. \ref{fig:pcon} that Dir-CRNs has higher value of $p^d_{con}$ than $p^{od}_{con}$ of Omn-Dir-CRNs and $p^o_{con}$ of Omn-CRNs; this implies that using directional antennas instead of omni-directional antennas in CRNs can significantly improve the connectivity of SUs. This improvement mainly owes to the \emph{higher spectrum availability} and the \emph{higher topological connectivity} as indicated in Section \ref{sec:connectivity}. In addition, aligning Fig. \ref{fig:pcon_r_a3} with Fig. \ref{fig:pcon_r_a5} together, we find that  $p^d_{con}$ is always higher than $p^{od}_{con}$ and $p^o_{con}$ in when $\alpha=3$ and $\alpha=5$.


\subsubsection{Impact of density of PUs $\lambda_p$}

In the second set of simulations, we investigate the impact of the density of PUs $\lambda_p$. The simulation results are shown in Fig. \ref{fig:pcon_lambda_p}. Similarly, we can observe that the simulation results (represented by markers) match with the analytical results (represented by curves), implying the accuracy of the proposed model. Meanwhile, we also find that the probability of connection of Dir-CRNs, Omn-Dir-CRNs and Omn-CRNs decreases with the increment of node density of PUs; this implies that \emph{the activities of PUs have a strong influence on the probability of connection of SUs}. Moreover, we also find that Dir-CRNs always outperform Omn-Dir-CRNs and Omn-CRNs in terms of the probability of connection when $\alpha=3$ and $\alpha=5$.
\begin{figure}[t]
\centering
\begin{tabular}{c c}
\begin{minipage}[t]{4.1cm}
\centering
\includegraphics[width=4.1cm]{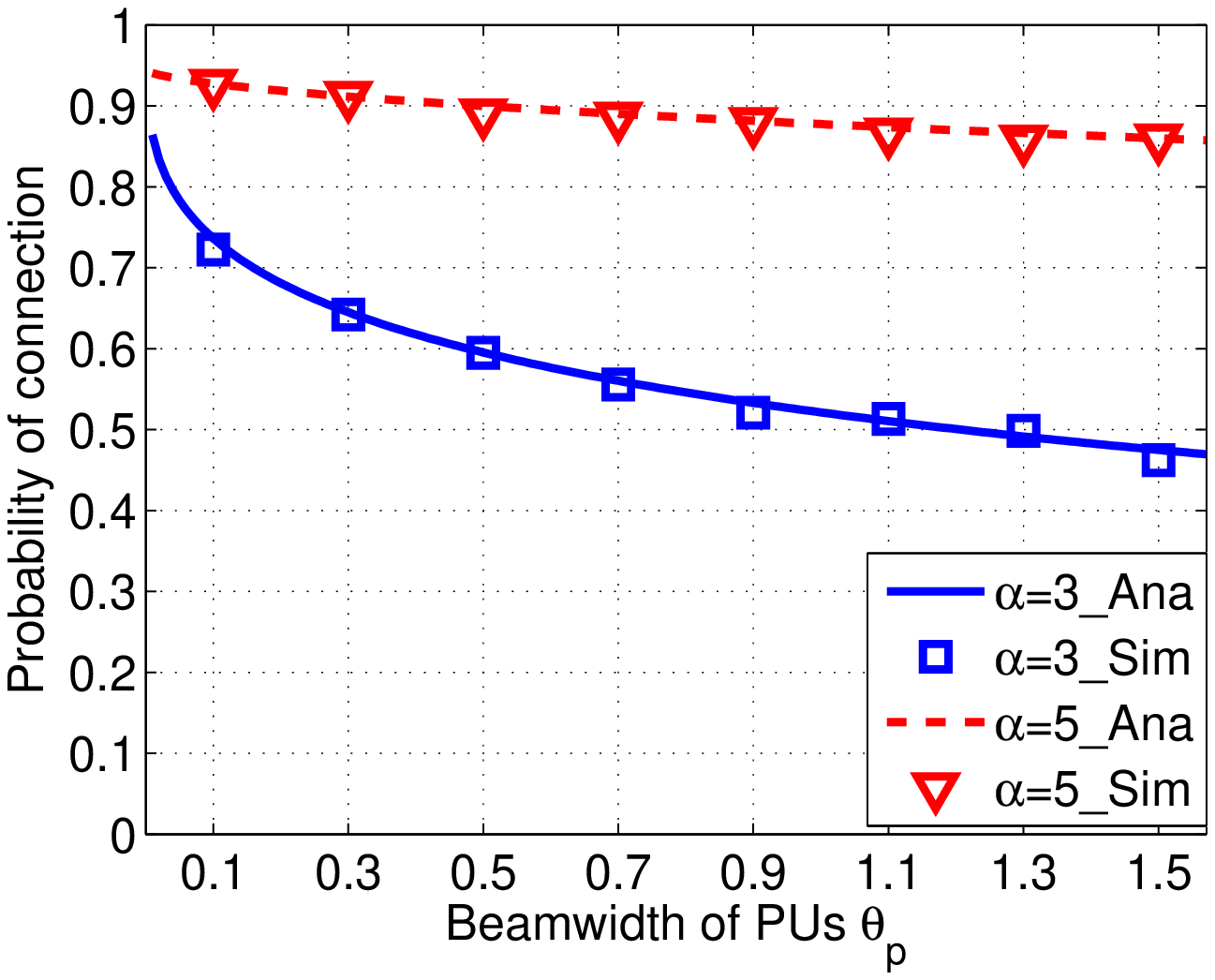}
\caption{Probability of connection of an SU pair versus the beamwidth of PUs, where $\lambda_p=0.01$, $r=3$ and $\theta_s=\pi/3$.}
\label{fig:pcon_theta_p}
 \end{minipage}
&
 \begin{minipage}[t]{4.1cm}
\centering
\includegraphics[width=4.1cm]{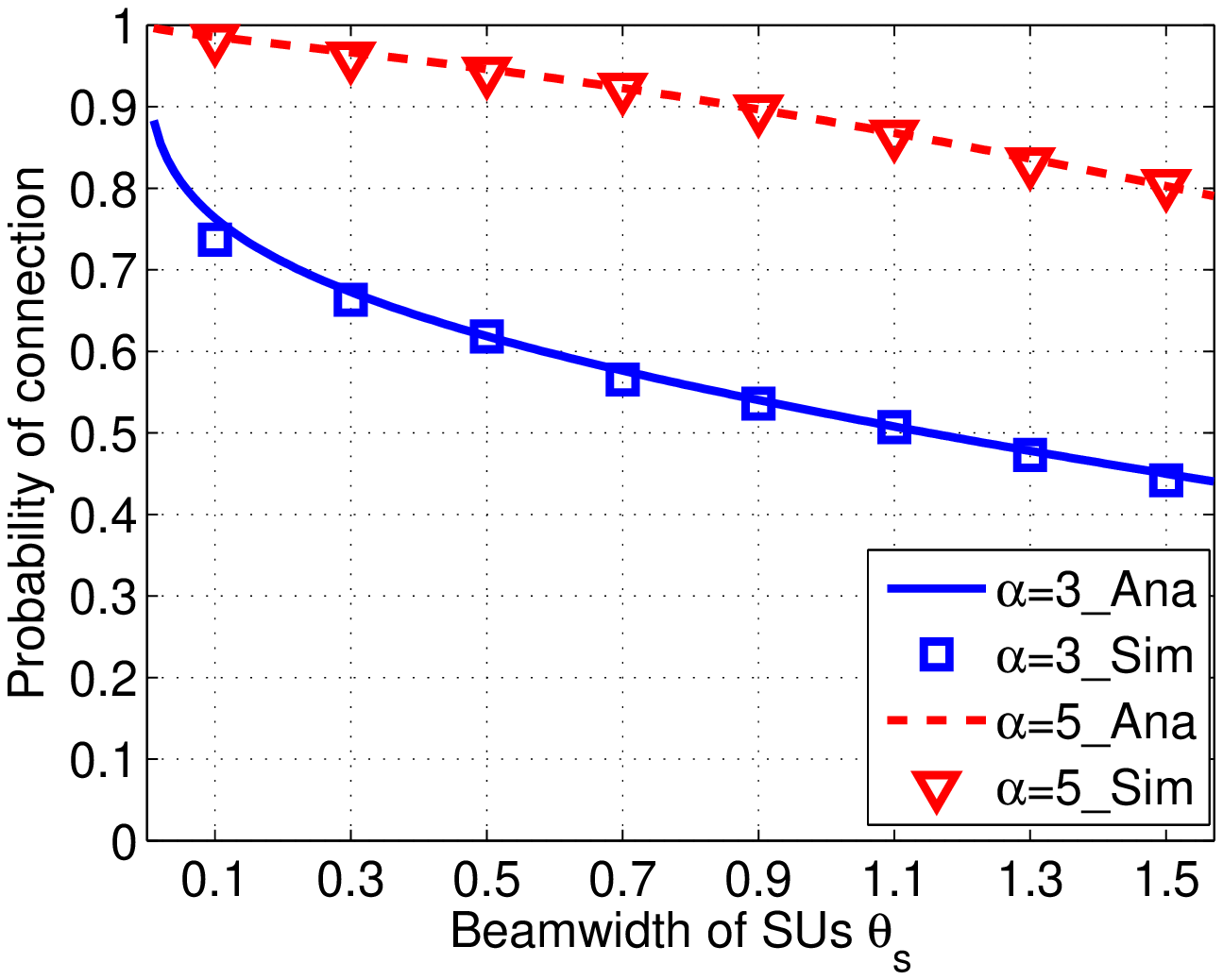}
\caption{Probability of connection of an SU pair versus the beamwidth of SUs, where $\lambda_p=0.01$, $r=3$ and $\theta_p=\pi/3$.}
\label{fig:pcon_theta_s}
 \end{minipage}
 \\
 \end{tabular}
\vspace*{-0.5cm}
\end{figure}

\subsubsection{Impact of beamwidth of PUs and SUs}

We next investigate the impact of different beamwidth of PUs and SUs on the probability of connection in Dir-CRNs. The third set of simulations are conducted to investigate the impact of the beamwith of PUs $\theta_p$. Fig. \ref{fig:pcon_theta_p} presents the results where we fix $\theta_s=\frac{\pi}{3}$ and vary $\theta_p$ in the range of $(0,\frac{\pi}{2}]$. We observe from Fig. \ref{fig:pcon_theta_p} that the probability of connection with larger path loss exponent $\alpha$ (e.g., $\alpha=5$) is higher than that with smaller path loss exponent $\alpha$ (e.g., $\alpha=3$). Moreover, increasing the beamwidth of PUs $\theta_p$ leads to the decreased probability of connection. This may owe to the effect that fewer SUs can have the spectrum since more SUs receive the detection preambles from PRs and the decreased topological connectivity of SUs due to the increased interference of PUs.

We further conduct the fourth set of simulations to investigate the impact of the beamwith of SUs $\theta_s$. Fig. \ref{fig:pcon_theta_s} presents the results where we fix $\theta_p=\frac{\pi}{3}$ and vary $\theta_s$ in the range of $(0,\frac{\pi}{2}]$. Fig. \ref{fig:pcon_theta_s} also shows that the probability of connection with larger path loss exponent $\alpha$ is higher than that with smaller path loss exponent $\alpha$ due to the higher spectrum availability and the higher topological connectivity brought by larger path loss exponent $\alpha$ (as indicated in Fig. \ref{fig:pij} and Fig. \ref{fig:ptop}). Meanwhile, increasing the beamwidth of SUs $\theta_s$ leads to the decreased probability of connection as shown in Fig. \ref{fig:pcon_theta_s}. This is because 1) more SUs receive the detection preambles from PRs due to the broader beamwidth of SUs and 2) the shorter transmission range of SUs due to the lower antenna gains.

%

\section{Conclusion}
\label{sec:conclusion}
In this paper, we analyse the connectivity of SUs in underlay cognitive radio networks with directional antennas (named as Dir-CRNs). Our model takes both the spectrum availability of SUs and the topological connectivity of SUs into account. Extensive simulation results verify the accuracy of our proposed model. In conclusion, this paper provides the following major findings:
\begin{itemize}
\item The connectivity of SUs heavily depends on the \emph{spectrum availability} and the \emph{topological connectivity}.
\item Dir-CRNs have the higher connectivity than Omn-CRNs. This improvement mainly owes to the lower interference caused by PUs, the higher spectrum availability and the higher topological connectivity, both of which are brought by directional antennas.
\item Dir-CRNs have the higher transmission power efficiency than Omn-CRNs. In particular, it requires less transmission power to establish a PU link or an SU link in Dir-CRNs than that in Omn-CRNs.
\item Dir-CRNs have the higher throughput of SUs than Omn-CRNs due to the lower interference, the higher spectrum availability and the higher topological connectivity.
\end{itemize}
Our proposed Dir-CRNs can potentially improve the performance of SUs in practical scenarios, such as smart grids \cite{YZhang:NetworkMag12,RYu:TII16}, vehicular ad hoc networks \cite{HeZLS:JCIN17} and mobile crowdsensing \cite{GYang:ComMag17,GYang:IOTJ17}. In addition, we also extend our analysis with consideration of transmission power efficiency and further investigate the throughput capacity of SUs (details can be found in \cite{DirCRN:2018}).

%


%
%

\tiny\bibliography{IEEEabrv,cognitiveradio}

\end{document}